\documentclass{svmult}

\usepackage{makeidx}         
\usepackage[]{graphicx}        
\usepackage{epstopdf}
\usepackage{amsmath}
\usepackage{amssymb}
\usepackage{amsfonts}

\makeindex             

\begin{document}

\title*{
A General Strategy for Physics-Based \\ Model Validation \\
Illustrated with Earthquake Phenomenology, \\ Atmospheric Radiative Transfer, \\
and Computational Fluid Dynamics
}
\titlerunning{A General Strategy for Physics-Based Model Validation}

\author{
Didier Sornette$^{1,2,3}$, Anthony B. Davis$^{4}$, James R. Kamm$^{5}$, and Kayo Ide$^{6}$}
\authorrunning{D. Sornette, A.B. Davis, J.R. Kamm, and K. Ide}

\institute{
$^1$ Institute of Geophysics and Planetary Physics \\ 
and 
Department of Earth \& Space Sciences \\
University of California, Los Angeles, CA 90095, USA. \\
$^2$ Laboratoire de Physique de la Mati\`ere Condens\'ee (CNRS UMR 6622) \\  
and
Universit\'e de Nice-Sophia Antipolis \\
06108 Nice Cedex 2, France \\
$^3$ now at D-MTEC, ETH Z\"urich, CH-8032 Z\"urich, Switzerland \\
\texttt{dsornette@ethz.ch} \\
\vskip .5cm
$^4$ Los Alamos National Laboratory \\ 
Space \& Remote Sensing Group (ISR-2) \\  
Los Alamos, NM 87545, USA \\
\texttt{adavis@lanl.gov} \\
\vskip .5cm
$^5$ Los Alamos National Laboratory \\  
Applied Science \& Methods Development Group (X-1) \\  
Los Alamos, NM 87545, USA \\
\texttt{kammj@lanl.gov} \\
\vskip .5cm
$^6$ Institute of Geophysics and Planetary Physics \\  
and
Department of Atmospheric \& Oceanic Sciences \\
University of California, Los Angeles, CA 90095, USA \\
\texttt{kayo@atmos.ucla.edu}
}
%
%
\maketitle

\vskip .5cm

This article is to be published the Lecture Notes in Computational Science and Engineering, Vol. TBD), Proceedings of in Computational Methods in Transport, Granlibakken 2006, 
F. Graziani and D. Swesty (Eds.),
Springer-Verlag, New York (NY), 2007.

\newpage
\noindent
This article is an augmented version of Ref. \cite{Sor07} by Sornette et al. that appeared in {\it Proceedings of the National Academy of Sciences} in 2007 (doi: 10.1073/pnas.0611677104), with an electronic supplement at URL \\ 
http://www.pnas.org/cgi/content/full/0611677104/DC1.  \\
Ref. \cite{Sor07}  is also available in preprint form at URL \\ 
http://arxiv.org/abs/physics/0511219.

\vskip 3cm

\abstract{
Validation is often defined as the process of determining the degree to which a model is an accurate representation of the real world from the perspective of its intended uses. Validation is crucial as industries and governments depend increasingly on predictions by computer models to justify their decisions. In this article, we survey the model validation literature and propose to formulate validation as an iterative construction process that mimics the process occurring implicitly in the minds of scientists. We thus offer a formal representation of the progressive build-up of trust in the model, and thereby replace incapacitating claims on the impossibility of validating a given model by an adaptive process of constructive approximation. This approach is better adapted to the fuzzy, coarse-grained nature of validation. Our procedure factors in the degree of redundancy versus novelty of the experiments used for validation as well as the degree to which the model predicts the observations. We illustrate the new methodology first with the maturation of Quantum Mechanics as the arguably best established physics theory and then with several concrete examples drawn from some of our primary scientific interests: a cellular automaton model for earthquakes,  an anomalous diffusion model for solar radiation transport in the cloudy atmosphere, and a computational fluid dynamics code for the Richtmyer--Meshkov instability.}

\newpage 

\section{Introduction: Our Position with Respect to Previous Work on Validation and Related Concepts}

\subsection{Introductory Remarks and Outline}

At the heart of the scientific endeavor, model building involves a slow and arduous selection process, which can be roughly represented as proceeding according to the following steps: 
\begin{enumerate}
\item
start from observations and/or experiments; 
\item
classify them according to regularities that they may exhibit: the presence of patterns, of some order, also sometimes referred to as structures or symmetries, is begging for ``explanations'' and is thus the nucleation point of modeling; 
\item
use inductive reasoning, intuition, analogies, and so on, to build hypotheses from which a model
\footnote{
By model, we understand an abstract conceptual construction based on axioms and logical relations developed to extract logical propositions and predictions.
}
is constructed; 
\item
test the model obtained in step 3 with available observations, and then extract predictions that are tested against new observations or by developing dedicated experiments. 
\end{enumerate}
The model is then rejected or refined by an iterative process, a loop going from step~1 to step~4. A given model is progressively validated by the accumulated confirmations of its predictions by repeated experimental and/or observational tests.

Building and using a model requires a language, i.e., a vocabulary and syntax, to express it. The language can be English or French for instance to obtain predicates specifying the properties of and/or relation with the subject(s). It can be mathematics, which is arguably the best language to formalize the relation between quantities, structures, space and change. It can be a computer language to implement a set of relations and instructions logically linked in a computer code to obtain quantitative outputs in the form of strings of numbers. In this later version, our primary interest here, validation must be distinguished from verification. Whereas {\it verification}\/ deals with whether the simulation code correctly solves the model equations, {\it validation}\/ carries an additional degree of trust in the value of the model vis-\`a-vis experiment and, therefore, may convince one to use its predictions to explore beyond known territories \cite{Roa98a}.

The validation of models is becoming a major issue as humans are increasingly faced with decisions involving complex tradeoffs in problems with large uncertainties, as for instance in attempts to control the growing anthropogenic burden on the planet within a risk-cost framework \cite{Cos97,Pim97} based on predictions of models. For policy decisions, national, regional, and local governments increasingly depend on computer models that are scrutinized by scientific agencies to attest to their legitimacy and reliability. Cognizance of this trend and its scientific implications is not lost on the engineering \cite{Bab04} and physics \cite{Pos05} communities.

Our purpose here is to clarify {\it from a physics-based perspective}\/ what validation is and to propose a roadmap for the development of systematic approach to physics-based validation with broad applications.  We will focus primarily on the needs of computational fluid dynamics and particle/radiation transport codes.

In the remainder of this section, we first review different definitions and approaches found in the literature, positioning ourselves with respect to selected topics or practices pertaining to validation; we then show how the validation problem is related to the mathematical statistics of hypothesis testing and discuss some problems associated with emergent behaviors in complex systems. In section~2, we list and describe qualitatively the elements required in our vision of model validation as an iterative process where one strives to build trust in the model going from one experiment to the next; however, one must also be prepared to uncover in the model a flaw, which may or may not be fatal. We offer in sections~3--4 our quantitative physics-based approach to model validation, where the relevance of the experiment to the validation process is represented explicitly. (An appendix explores the model validation problem more formally and in a broader context.) Section~5 demonstrates the general strategy for model validation using the historical development of quantum physics---a remarkably clear ideal case. Section~6 uses some research interests of the present authors to further illustrate the validation procedure using less-than-perfect models in geophysics,  computational fluid dynamics (CFD), and radiative transfer.  We summarize in section~7.

\subsection{Standardized Definitions}

\begin{figure}
\begin{center}
\includegraphics[width=6cm]{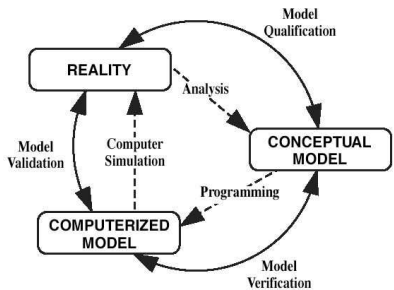}
\caption{Schematic representation of the conventional position of validation in model construction according to Schlesinger \cite{Sch79} and Sargent  \cite{Sar98,Sar01}.}
\label{f:VV}
\end{center}
\end{figure}

The following definitions are given by the 
American Institute of Aeronautics and Astronautics \cite{AIA98}:
\begin{itemize}
\item {\it Model}: 
A representation of a physical system or process intended to enhance our ability to predict, control and eventually to understand its behavior.
\item {\it Calibration}:
The process of adjusting numerical or physical modeling parameters in the computational model for the purpose of improving agreement with experimental data.
\item {\it Verification}: 
The process of determining that a model implementation accurately represents the developer's conceptual description of the model and the solution of the model.
\item {\it Validation}: 
The process of determining the degree to which a model is an accurate representation of the real world from the perspective of the intended uses of the model.
\end{itemize}
Figure \ref{f:VV}, sometimes called a Sargent diagram, shows where validation and several other of the above constructs and stages enter into a complete modeling project.

In the concise phasing of Roache \cite{Roa98a}, {\it ``Verification consists in solving the equations right while validation is solving the right equations.''} In the context of the validation of astrophysical simulation codes, Calder et al. \cite{Cal02} add: {\it ``Verification and validation are fundamental steps in developing any new technology. For simulation technology, the goal of these testing steps is assessing the credibility of modeling and simulation.''}

{\it Verifications}\/ of complex CFD codes usually comprise a suite of standard test problems in the field of fluid dynamics \cite{Cal02}. These include Sod's test \cite{Sod78}, the strong shock tube problem \cite{Rid99}, the Sedov explosion problem \cite{Sed59}, the interacting blast wave problem \cite{Woo84}, a shock forced through a jump in mesh refinement, and so on.

{\it Validations}\/ of complex CFD codes is usually done by comparison with experiments testing a variety of physical phenomena, including instabilities, turbulent mixing, shocks, etc. Validation requires that the numerical simulations recover the salient qualitative features of the experiments, such as the instabilities, their nonlinear development, the determination of the most unstable modes, and so on. See, for instance, Gnoffo et al. \cite{Gno98}.

Considerable work on verification and validation of simulations has been done in the field of CFD, and in this literature the terms verification and validation have precise, technical meanings \cite{AIA98,Roa98a,Roa98b,Sar98,Sar01}. Verification is taken to mean demonstrating that a code or simulation accurately represents the conceptual model. Roache~\cite{Roa02} stresses the importance of distinguishing between (i) verification of codes and (ii) verification of calculations. The former is concerned with the correctness of the code. The later deals with the correctness of the physical equations used in the code. The programming and methods of solution can be correct (verification (i) successful) but they can solve erroneous equations (verification (ii) failure). Validation of a simulation means demonstrating that the simulation appropriately describes Nature. The scope of validation is therefore much larger than that of verification and includes comparison of numerical results with experimental or observational data. In astrophysics, where it is difficult to obtain observations suitable for comparison to numerical simulations, this process can present unique challenges. Roache [op. cit.] goes on to offer the optimistic prognosis that {\it ``the problems of Verification of Codes and Verification of Calculations are essentially solved for the case of structured grids, and for structured refinement of unstructured grids. It would appear that one higher level of algorithm/code development is required in order to claim a complete methodology for Verification of Codes and Calculations. I expect this to happen. Within 10 years, and likely much less, Verification of Codes and Calculations ought to be settled questions. I expect that Validation questions will always be with us.''} We fully endorse this last sentence, as we will argue further on that validation is akin to the development of ``trust'' in theories of real phenomena, a never-ending quest.

\subsection{Impossibility Statements}

For these reasons, the possibility of validating numerical models of natural phenomena, often endorsed either implicitly or identified as reachable goals by natural scientists in their daily work, has been challenged; quoting from Oreskes et al. \cite{Ore94a}: {\it ``Verification and validation of numerical models of natural systems is impossible. This is because natural systems are never closed and because model results are always non-unique.''} According to this view, the impossibility of ``verifying'' or ``validating'' models is not limited to computer models and codes but to all theories that rely necessarily on imperfectly measured data and auxiliary hypotheses.  As Sterman \cite{Ste94} puts it: {\it ``Any theory is underdetermined and thus unverifiable, whether it is embodied in a large-scale computer model or consists of the simplest equations.''} Accordingly, many uncertainties undermine the predictive reliability of any model of a complex natural system in advance of its actual use.
\footnote{
For further debate and commentary by Oreskes and her co-authors, see refs. \cite{Ryk94,Ore94b,Ore98}; also noteworthy is the earlier paper by Konikov and Bredehoeft \cite{Kon92} for a statement about validation impossibility in the context of groundwater models.
}

Such ``impossibility'' statements are reminiscent of other ``impossibility theorems.'' Consider the mathematics of algorithmic complexity \cite{Cha87}, which provides one approach to the study of complex systems. Following reasoning related to that underpinning G\"odel's incompleteness theorem, most complex systems have been proved to be computationally irreducible, i.e., the only way to predict their evolution is to actually let them evolve in time. Accordingly, the future time evolution of most complex systems appears inherently unpredictable. Such sweeping statements turn out to have basically no practical value. This is because, in physics and other related sciences, one aims at predicting coarse-grained properties. Only by ignoring most of molecular detail, for example, did researchers ever develop the laws of thermodynamics, fluid dynamics and chemistry. Physics works and is not hampered by computational irreducibility because we only ask for approximate answers at some coarse-grained level \cite{Buc05}. By developing exact but {\it coarse-grained} procedures on computationally irreducible cellular automata, Israeli and Goldenfeld \cite{Isr04} have demonstrated that prediction may simply depend on finding the right level for describing the system. More generally, we argue that only coarse-grained scales are of interest in practice but their description requires ``effective'' laws which are in general based on finer scales. In other words, real understanding must be rooted in the ability to predict coarser scales from finer scales, i.e., a real understanding solves the universal micro-macro challenge. Similarly, we propose that validation is possible, to some degree, as explained further on.

\subsection{Validation and the Mathematical Statistics of Hypothesis Testing}

Calder et al. \cite{Cal02} also write: {\it ``We note that verification and validation are necessary but not sufficient tests for determining whether a code is working properly or a modeling effort is successful. These tests can only determine for certain that a code is not working properly.''} This last statement is important because it points to a bridge between the problem of validation and some of the most central questions of mathematical statistics \cite{Bor98}, namely, hypothesis testing and statistical significance tests. This connection has been made previously by several others authors \cite{Col97,Hil99,Hil00,Eas01}. In showing the usefulness of the concepts and framework of hypothesis testing, we depart from Oberkampf and Trucano \cite{Obe02a} who mistakenly state that hypothesis testing is a true or false issue, only. Every test of significance begins with a ``null'' hypothesis H$_0$, which represents a theory that has been put forward, either because it is believed to be true or because it is to be used, but has not been proved.
\footnote{
We refer the reader to V.J. Easton and J.H. McColl, {\it Statistics Glossary},
http://www.cas.lancs.ac.uk/glossary\_v1.1/main.html, 
from which we have borrowed liberally for this brief summary.
}

For example, in a clinical trial of a new drug, the null hypothesis might be: ``the new drug is no better, on average, than the current drug.'' We would write H$_0$: ``there is no difference between the two drugs on average.'' The alternative hypothesis H$_1$ is a statement of what a statistical hypothesis test is set up to establish. In the example of a clinical trial of a new drug, the alternative hypothesis might be that the new drug has a different effect, on average, to be compared to that of the current drug. We would write H$_1$: the two drugs have different effects, on average. The alternative hypothesis might also be that the new drug is better, on average, than the current drug. Once the test has been carried out, the final conclusion is always given in terms of the null hypothesis. We either ``reject H$_0$ in favor of H$_1$'' or ``do not reject H$_0$.'' We never conclude ``reject H$_1$,'' or even ``accept H$_1$.'' If we conclude ``do not reject H$_0$,'' this does not necessarily mean that the null hypothesis is true, it only suggests that there is not sufficient evidence against H$_0$ in favor of H$_1$; rejecting the null hypothesis then suggests that the alternative hypothesis may be true, or is at least better supported by the data. Thus, one can never prove that an hypothesis is true, only that it is wrong by comparing it with another hypothesis. One can also conclude that ``hypothesis H$_1$ is not necessary and another, more parsimonious, one $H_0$ should be favored.'' The alternative hypothesis H$_1$ is not rejected, strictly speaking, but is found unnecessary or redundant with respect to H$_0$. This is the situation when there are two (or several) alternative hypotheses H$_0$ and H$_1$, which can be composite, nested, or non-nested.
\footnote{
The technical difficulties of hypothesis testing depend on these nested structures of the competing hypotheses; see, for instance, Gourieroux and Monfort \cite{Gor94}.
}

Within this framework, the above-mentioned statement by Oreskes et al. \cite{Ore94a} that verification and validation of numerical models of natural systems is impossible is hardly news: the theory of statistical hypothesis testing has taught mathematical and applied statisticians for decades that one can never prove an hypothesis or a model to be true. One can only develop an increasing trust in it by subjecting it to more and more tests that ``do not reject it.'' We attempt to formalize below how such trust can be increased to lead to an asymptotic validation.

\subsection{Code Comparison}

The above definitions are useful in recasting the role of code comparison in verification and validation (Code Comparison Principle or CCP). Trucano et al. \cite{Tru03} are unequivocal on this practice: {\it ``the use of code comparisons for validation is improper and dangerous.''} We propose to interpret the meaning of CCP for code verification activities (which has been proposed in this literature) as parallel to the problem of hypothesis testing: Can one reject Code \#1 in favor of Code \#2? In this spirit, the CCP is nothing but a reformulation in the present context of the fundamental principle of hypothesis testing. Viewed in this way, it is clear why CCP is not sufficient for validation since validation requires comparison with experiments and several other steps described below. The analogy with hypothesis testing illuminates what CCP actually is: CCP allows the selection of one code among several codes (at least two) but does not help one to draw conclusions about the validity of a given code or model when considered as a unique entity independent of other codes or models.
~\footnote{
We should stress that the Sandia Report \cite{Tru03} by Trucano et al. presents an even more negative view of code comparisons because it addresses the common practice in the simulation community that turns to code comparisons rather than bone fide verification or validation, without any independent referents.
}
Thus, the fundamental problem of validation is more closely associated with the other class of problems addressed by the theory of hypothesis testing, which consists in the so-called ``tests of significance'' where one considers only a single hypothesis H$_0$, and the alternative is ``all the rest,'' i.e., all hypotheses that differ from H$_0$. In that case, the conclusion of a test can be the following: ``this data sample does not contradict the hypothesis H$_0$,'' which is not the same as ``the hypothesis H$_0$ is true.'' In other words, an hypothesis cannot be excluded because it is found sufficient at some confidence level for explaining the available data. This is not to say that the hypothesis is true. It is just that the available data is unable to reject said hypothesis. Restating the same thing in a positive way, the result of a test of significance is that the hypothesis H$_0$ is ``compatible with the available data.''

It is implicit in the above discussion that, to compare codes quantitatively in a meaningful way, they must solve the same set of equations using different algorithms, and not just model the same physical system.  Indeed, there is nothing wrong with ``validating'' a numerical implementation of a knowingly approximate approach to a given physical problem.  For instance, a (duly verified) diffusion/P$_1$ transport code can be validated against a detailed Monte Carlo or S$_n$ code. The more detailed model must in principle be validated against real-world data. In turn, it provides validation ``data'' to the coarser model. Naturally, the coarser (say, P$_1$ transport) model still needs to establish its relevance to the real world problem of interest, preferably by comparison with real observations, or at least be invoked only in regimes where it is known a priori to be sufficiently accurate based on comparison with a finer (say, Monte Carlo transport) model.

Two noteworthy initiatives in transport model comparison for non-nuclear applications are the Intercomparison of 3D Radation Codes (I3RC) \cite{Cah05} (i3rc.gsfc.nasa.gov) and the RAdiation Model Intercomparison (RAMI) \cite{Pin01,Pin04} (rami-benchmark.jrc.it).  The former is focused on the challenge of 3D radiative transfer in the cloudy atmosphere while the later is about 3D radiative transfer inside plant canopies; both efforts are motivated by issues in remote sensing (especially from space) and radiative energy budget estimation (either in the framework of climate modeling or using observational diagnostics, which typically means more remote sensing).
\footnote{
In remote sensing science, transport theory (for photons) plays a central role and ``validation'' has a special meaning, namely, the estimation of uncertainty for remote sensing products based on ``ground-truth,'' i.e., field measurements of the very same geophysical variables (e.g., surface temperature or reflectivity, vegetation productivity, soil moisture) that the satellite instrument is designed to quantify. These data are collected at the same location as the imagery, if possible, at the precision of a single pixel.  This type of validation exercise will test both the ``forward'' radiation transport theory and its ``inversion.'' Atmospheric remote sensing, particularly of clouds, poses a special challenge because, strictly-speaking, there is no counterpart of ground-truthing. One must therefore often make do with comparisons of ground-based and space-based remote-sensing (say, of the column-integrated aerosol burden) to quantify uncertainty in both operations.  In-situ measurements (temperature, humidity, cloud liquid water, etc.) from airborne platforms---balloon or aircraft---are always welcome but collocation is rarely close enough for point-to-point comparisons; statistical agreement is then all that is to be expected, and residuals provide the required uncertainty.
}
Much has been learned by the modelers participating in these code comparison studies, and the models have been improved on average \cite{Wid07}. Although not connected so far to the engineering community that is at the forefront of V\&V standardization and methodology, the I3RC and RAMI communities talk much about ``testing,'' and sometimes ``certification,'' and not so much about ``verification'' (which would be appropriate) or ``validation'' (which would not).

What about multi-physics codes such as those used routinely in astrophysics, nuclear engineering, or climate modeling?  CCP, along with the stern warnings of Trucano et al. \cite{Tru03}, applies here, too.  Even assuming that all the model components are properly verified or even individually validated, the aggregated model is likely to be too complex to talk about clean verification through output comparison. Finding some level of agreement between two or more complex multi-physics models will naturally build confidence in the whole (community-wide) modeling enterprise. However, this is not to be interpreted as validation of any or all of the individual models.

There are many reasons for wanting to have not just one model on hand but a suite of more or less elaborate ones.  A typical collection can range from the mathematically and physically exact but numerically intractable to the analytically solvable, possibly even on the proverbial back-of-an-envelope. We elaborate on and illustrate this kind of hierarchical modeling effort in section \ref{s:4pbs} of the Appendix, offering it as an approach where model development is basically simultaneous with its validation.

\subsection{Relations Between Validation, Calibration and Data Assimilation}
\label{s:cal+DA}

As previously stated, validation can be characterized as the act of quantifying the credibility of a model to represent phenomena of interest. Virtually all such models contain numerical parameters, the precise values of which are not known a priori and, therefore, must be assigned.  Calibration is the process of adjusting those parameters to optimize (in some sense) the agreement between the model results and a specific set of experimental data.  Such data necessarily have uncertainties associated with them, e.g., due to natural variability in physical phenomena as well as to unavoidable imprecision of diagnostics. Likewise, there are intrinsic errors associated with the numerical methods used to evaluate many models, e.g., in the approximate solutions obtained from discretization schemes applied to partial differential equations.  The approach of defensibly prescribing parameters for complex physical phenomena while incorporating the inescapable variability in these values is called ``calibration under uncertainty,'' \cite{Tru06} a field that poses non-trivial challenges in its own right.

However calibration is approached, it must be undertaken using a set of data---ideally from specifically chosen calibration experiments/observations \cite{Obe02b}---that differs from the physical configurations of ultimate interest (i.e., against which the model will be validated).  In order to ensure that validation remains independent of calibration, it is imperative that these data sets be disjoint.  In the case of large, complex, and costly experiments encountered in many real-world applications, it can be difficult to maintain a scientific ``demilitarized zone'' between calibration and validation. To not do so, however, risks undermining the scientific integrity of the associated modeling enterprise, the potential predictive power of which may rapidly wither as the validation study devolves into a thinly disguised exercise in calibration.

For complex systems, there are many choices to be made regarding experimental and numerical studies in both validation and calibration. The high-level approach of the Phenomena Identification and Ranking Table (PIRT) \cite{Tru02} can be used to heuristically characterize the nature of one's interest in complicated systems. This approach uses expert knowledge to identify the phenomenological components in a system of interest, to rank their (relative) perceived importance in the overall system, and to gauge the (relative) degree to which these component phenomena are perceived to be understood.  This rough-and-ready approach can be used to target the choice of validation experiments for the greatest scientific payoff on fixed experimental and simulation budgets.  To help guide calibration activities, one can apply the quantitative techniques of sensitivity analysis to rank the relative impact of input parameters on model outcome.  Such considerations are particularly important for complex models containing many adjustable parameters, for which it may prove impossible to faithfully calibrate all input parameters.

Saltelli et al. \cite{Sal00, Sal04} have championed ``sensitivity analysis'' methods, which come in two basic flavors and many variations. One class of methods uses exact or numerical evaluation of partial derivatives of model output deemed important with respect to input parameters to seek regions of parameter space that might need closer examination from the standpoints of calibration and/or validation.  If the model has time dependence, one can follow the evolution of how parameter choices influence the outcome.  The alternate methodology uses adjoint dynamical equations to determine the relative importance of various parameters. The publications of Saltelli et al. provide numerous examples illustrating the value and practical impact of sensitivity analysis, as well as references to the wide scientific literature on this subject.  The results of numerical studies guided by sensitivity analysis can be used both to focus experimental resources on high-impact experimental studies and to steer future model development efforts.

In dynamical modeling, initial conditions can be viewed as parameters and, as such, they need to be determined optimally from data. If the dynamical system in question is evolving continuously over time and data become available along the trajectory of the dynamical system, the problem of finding {\it a single} initial condition over the entire trajectory becomes increasingly and exceedingly difficult as the time window of the trajectory extends. In fact, it is practically impossible for the systems like the atmosphere or ocean whose dynamics is highly nonlinear, high-dimensional model is undoubtedly imperfect, and inhomogeneous and sporadic data are subject to (poorly understood) errors.

Data assimilation is an approach that attends to this problem by breaking up the trajectory  over (fixed-length) time windows and solving the initialization problem sequentially over one time window at a time as data become available. A novelty of data assimilation is that, rather than solving the initialization problem from scratch, it uses the model forecast as the first guess (the prior) of the initialization (optimization) problem. Once the optimization is completed, the optimal solution (the posterior) becomes the initial condition for the next model forecast.

This iterative Bayesian approach to data assimilation is most effective when the uncertainties in both the prior and the data are accurately quantified, as the system evolves over time and the data assimilation iterates one cycle after another. This is a non-trivial problem, because it requires the estimate of not only the model state but also the uncertainties associated with it, as well as the proper description of the uncertainties in data.

Numerical weather prediction (NWP) is one of the most familiar application areas of data assimilation---one with major societal impact. The considerable progress in skill of the NWP in recent decades has been due to improvements in all aspects of data assimilation  \cite{Kal03}, i.e., modeling of the atmosphere, quality and quantity of data, and data assimilation methods. At the time of writing, most operational NWP centers use the so-called the ``three-dimensional variational method'' (3D-Var) \cite{Par92}, which is an economical and accurate statistical interpolation scheme that does not include the effect of uncertainty in the forecast. Some centers have switched to the ``four-dimensional variational method'' (4D-Var) \cite{Rab00}, which incorporates the evolution of uncertainty in linear sense by the used of the adjoint model of the highly nonlinear model. These variational methods always call for the minimization of a cost function (cf. Appendix) that measures the difference between model results and observations throughout some relevant region of space and time. Currently active research areas in data assimilation include the effective and efficient quantification of the time-dependent uncertainties of both the prior and posterior in the analysis. To this end, the ensemble Kalman filter methods have recently received considerable attention motivated by future integration into operational environments \cite{Hou05,Szu07,Whi07}. As the importance of the uncertainties in data assimilation have become clear, many NWP centers perform ensemble prediction along with the single analysis obtained by the variational methods \cite{Tot93,Hou95,Mol96}.

Clearly, considerable similarities exist between the data assimilation problem and the model validation problem. Can successful data assimilation be construed as validation of the model?  In our opinion, that would be unjustified because the objectives are clearly different for these problems. As stated above, data assimilation admits the imperfection of the model. It explicitly makes use of the knowledge from the previous data assimilation cycle.
As the initialization problem is solved iteratively over relatively short time windows, deviation of the model trajectory from the true evolution of the dynamical system in question tend to be small and data could be assimilated into the model without much discrepancy. Moreover, the operational centers perform careful quality-control of data to eliminate any  isolated ``outliers'' with respect to the model trajectory. Thus, the data assimilation problem differs from the validation problem by design. Nevertheless, it is important to recognize that the resources offered by data assimilation can ensure that models perform well enough for their intended use.

\subsection{Extension of the Meaning of Validation}

A qualitatively new class of problems arise in fields such as the geosciences that deal with the construction of knowledge of a unique object, planet Earth, whose full scope and range of processes can be replicated or controlled neither in the laboratory nor in a supercomputer. This has led recently to championing the relevance of ``systemic'' (meaning ``system approach'') also called ``complex system'' approaches to the geosciences. In this framework, positive and negative feedbacks (and even more complicated nonlinear multiplicative noise processes) entangle many different mechanisms, whose impact on the overall organization can be neither assessed nor understood in isolation. How does one validate a model using the systemic approach? This very interesting and difficult question is at the core of the problem of validation.

How does one validate a model when it is making predictions on objects that are not fully replicated in the laboratory, either in the range of variables, of parameters, or of scales? For instance, this question is crucial 
\begin{itemize}
\item
in the scaling the physics of material and rock rupture tested in the laboratory to the scale of earthquakes;
\item
in the scaling the knowledge of hydrodynamical processes quantified in the laboratory to the length and time scales relevant to the atmospheric/oceanic weather and climate, not to mention astrophysical systems;
\item
in the science-based stewardship of the nuclear arsenal, where the challenge is to go from many component models tested at small scales in the laboratory to the full-scale explosion of an aging nuclear weapon.
\end{itemize}

\begin{figure}
\begin{center}
\includegraphics[width=12cm]{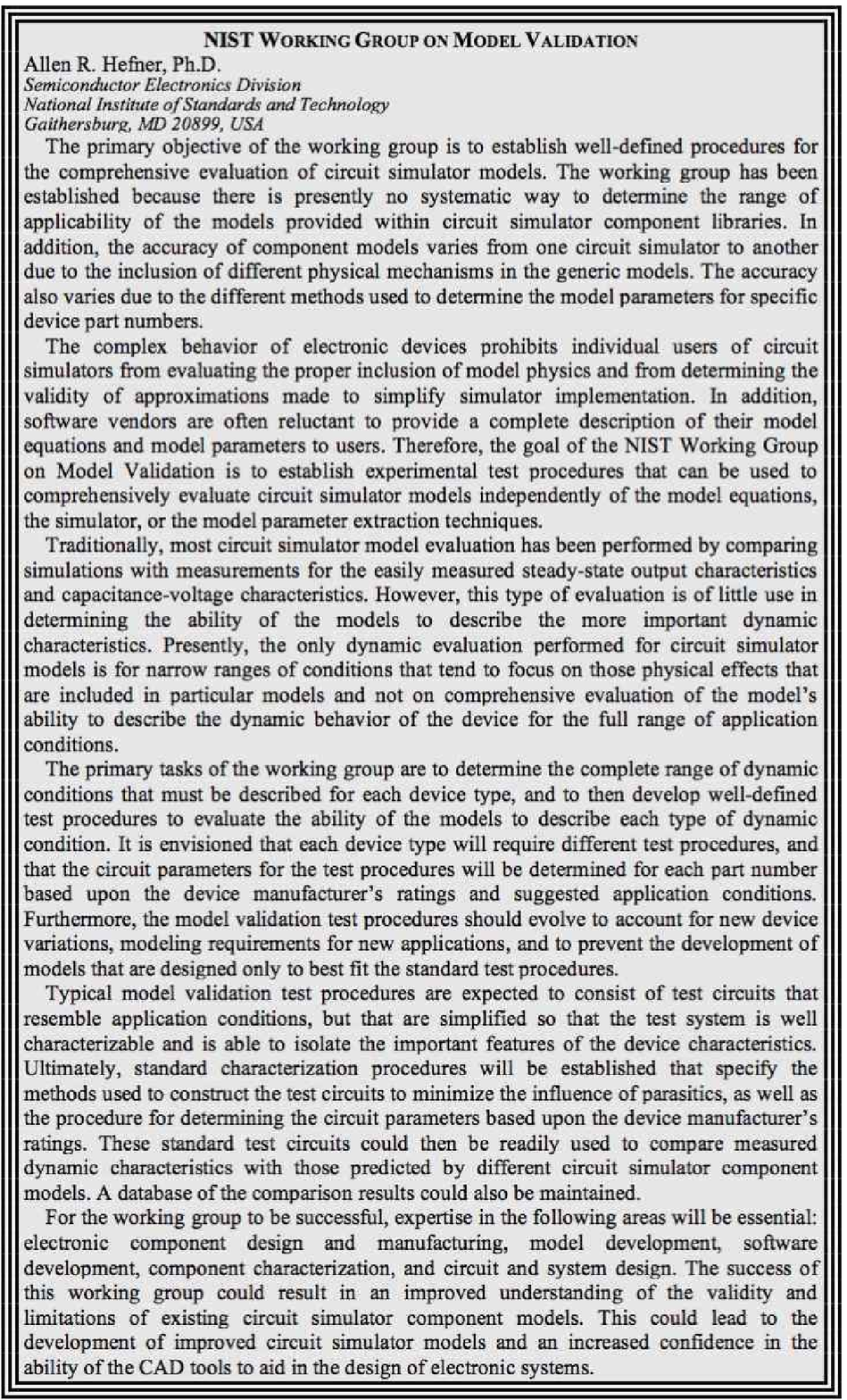}
\end{center}
\end{figure}

The same issue arises in the evaluation of electronic circuits. In 2003, Allen R. Hefner, Founder and Chairman of the NIST/IEEE Working Group on Model Validation, writes in its description: {\it ``The problem is that there is no systematic way to determine the range of applicability of the models provided within circuit simulator component libraries.''} See full-page boxed text for the complete version of this interesting text, as well as Ref.~\cite{Ber98}. This example of validation of electronic circuits is particularly interesting because it stresses the origin of the difficulties inherent in validation: the fact that the dynamics are nonlinear and complex with threshold effects and does not allow for a simple-minded analytic approach consisting in testing a circuit component by component. Extrapolating, this same difficulty is found in validating general circulation models of the Earth's climate or computer codes of nuclear explosions. The problem is thus fundamentally a ``system'' problem. The theory of systems, sometimes referred to as the theory of complex systems, is still in its infancy but has shown the existence of surprises. The biggest surprise may be the phenomenon of ``emergence'' in which qualitatively new processes or phenomena appear in the collective behavior of the system, while they cannot be derived or guessed from the behavior of each element. The phenomenon of ``emergence'' is similar to the philosophical law on the ``transfer of the quantity into the quality.'' How does one validate a model of such a system? Validation therefore requires an understanding of this emergence phenomenon.

From another angle, the problem is that of extrapolating a body of knowledge, which is firmly established only in some limited ranges of variables, parameters and scales, beyond this clear domain into a more fuzzy zone of unknowns. This problem has appeared and appears again and again in different guises in practically all scientific fields. A particularly notable domain of application is risk assessment; see, for instance, Kaplan and Garrick's classic paper on risks \cite{Kap81}, and the instructive history of quantitative risk analysis in US regulatory practice \cite{Rec00}, especially in the US nuclear power industry \cite{Kee91,Hel94,Hel99,Hel00}. An acute  question in risk assessment deals with the question of quantifying the potential for a catastrophic event (earthquake, tornado, hurricane, flood, huge solar mass ejection, large meteorite, industrial plant explosion, ecological disaster, financial crash, economic collapse, etc.) of amplitude never yet sampled from the knowledge of past history and present understanding.

To tackle this enduring question, each discipline has developed its own strategies, often being unaware of the approaches of others. Here, we attempt a formulation of the problem, and outline some general directions of attack, that hopefully will transcend the specificities of each discipline. Our goal is to formulate the validation problem in a way that may encourage productive crossings of disciplinary lines between different fields by recognizing the commonalities of the blocking points, and suggest useful guidelines.

\section{Validation as a Constructive Iterative Process}

In a generic exercise in model validation, one performs an experiment and, in parallel, runs the calculations with the available model. A comparison between the measurements of the experiment and the outputs of the model calculations is then performed. This comparison uses some metrics controlled by experimental feasibility, i.e., what can actually be measured. One then iterates by refining the model until (admittedly subjective) satisfactory agreement is obtained. Then, another set of experiments is performed, which is compared with the corresponding predictions of the model. If the agreement is still satisfactory without modifying the model, this is considered progress in the validation of the model. Iterating with experiments testing different features of the model corresponds to mimicking the process of construction of a theory in physics \cite{Dys88}. As the model is exposed to increasing scrutiny and testing, the testers develop a better understanding of the reliability (and limitations) of the model in predicting the outcome of new experimental and/or observational set-ups. This implies that {\it ``validation activity should be organized like a project, with goals and requirements, a plan, resources, a schedule, and a documented record''} \cite{Pos05}.

Extending previous work \cite{Col97,Hil99,Hil00,Eas01}, we thus propose to formulate the validation problem of a given model as an iterative construction that embodies the often implicit process occurring in the minds of scientists:

\begin{enumerate}

\item
One starts with an a priori trust quantified by the value $V_{\rm prior}$ in the potential value of the model. This quantity captures the accumulated evidence thus far. If the model is new or the validation process is just starting, take $V_{\rm prior} = 1$.  As we will soon see, the absolute value of $V_{\rm prior}$ is unimportant but its relative change is important.

\item
An experiment is performed, the model is set-up to calculate what should be the outcome of the experiment, and the comparison between these predictions and the actual measurements is made either in model space or in observation space. The comparison requires a choice of metrics.

\item 
Ideally, the quality of the comparison between predictions and observations is formulated as a statistical test of significance in which an hypothesis (the model) is tested against the alternative, which is ``all the rest.'' Then, the formulation of the comparison will be either ``the model is rejected'' (it is not compatible with the data) or ``the model is compatible with the data.'' In order to implement this statistical test, one needs to attribute a likelihood $p(M | y_{\rm obs})$ or, more generally, a metric-based ``grade'' that quantifies the quality of the comparison between the predictions of the model $M$ and observations $y_{\rm obs}$. This grade is compared with the reference likelihood $q$ of ``all the rest.'' Examples of implementations include the sign test and the tolerance interval methods. 
\footnote{
Pal and Makai \cite{Pal03} have used the mathematical statistics of hypothesis testing as a way to validate the correctness of code simulating the operation of a complex system with respect to a level of confidence for safety problems. The main conclusion is that the testing of the input variables separately may lead to incorrect safety related decisions with unforeseen consequences. They have used two statistical methods: the sign test and the tolerance interval methods for testing more than one mutually dependent output variables. We propose to use these and similar tests delivering a probability level $p$ which can then be compared with a pre-defined likelihood level $q$.
}
In many cases, one does not have the luxury of a likelihood; one has then to resort to more empirical assessments of how well the model explains crucial observations. In the most complex cases, the outcome can be binary (accepted or rejected).

\item
The posterior value of the model is obtained according to a formula of the type 
\begin{equation}
V_{\rm posterior}/V_{\rm prior}  =
F\left[ p(M | y_{\rm obs}), q; c_{\rm novel} \right]~.
\label{mgler}
\end{equation}
In this expression, $V_{\rm posterior}$ is the posterior potential, or coefficient, of trust in the value of the model after the comparison between the prediction of the model and the new observations have been performed. By the action of $F[\cdots]$, $V_{\rm posterior}$ can be either larger or smaller than $V_{\rm prior}$: in the former case, the experimental test has increased our trust in the validity of the model; in the later case, the experimental test has signaled problems with the model. One could call $V_{\rm prior}$ and $V_{\rm posterior}$ the evolving ``potential value of our trust'' in the model or, loosely paraphrasing the theory of decision making in economics, the ``utility'' of the model \cite{Neu44}.

\end{enumerate}

\noindent
The transformation from the potential value $V_{\rm prior}$ of the model before the experimental test to $V_{\rm posterior}$ after the test is embodied into the multiplier $F$, which can be either larger than $1$ (towards validation) or smaller than $1$ (towards invalidation). We postulate that $F$ depends on the grade $p(M | y_{\rm obs})$, to be interpreted as proportional to the probability of the model $M$ given the data $y_{\rm obs}$.  It is natural to compare this probability with the reference likelihood $q$ that one or more of all other conceivable models is compatible with the same data.

Our multiplier $F$ depends also on a parameter $c_{\rm novel}$ that quantifies the importance of the test. In other words, $c_{\rm novel}$ is a measure of the impact of the experiment or of the observation, that is, how well the new observation explores novel ``dimensions'' of the parameter and variable spaces of both the process and the model that can reveal potential flaws. A fundamental challenge is that the determination of $c_{\rm novel}$ requires, in some sense, a pre-existing understanding of the physical processes so that the value of a new experiment can be fully appreciated. In concrete situations, one has only a limited understanding of the physical processes and the value of a new observation is only assessed after a long learning phase, after comparison with other observations and experiments, as well as after comparison with the model, making $c_{\rm novel}$ possibly self-referencing. Thus, we consider $c_{\rm novel}$ as a judgment-based weighting of experimental referents, in which judgment (for example, by a subject matter expert) is dominant in its determination. The fundamental problem is to quantify the relevance of a new experimental referent for validation to a given decision-making problem, given that the experimental domain of the test does not overlap with the application domain of the decision. Assignment of $c_{\rm novel}$ requires the judgment of subject matter experts, whose opinions will likely vary. This variability must be acknowledged (if not accounted for, however naively) in assigning $c_{\rm novel}$. Thus, providing an a priori value for $c_{\rm novel}$, as required in expression (\ref{mgler}), remains a difficult and key step in the validation process. This difficulty is similar to specifying the utility function in decision making \cite{Neu44}.

Repeating an experiment twice is a special degenerate case since it amounts ideally to increasing the size of the statistical sample. In such a situation, one should aggregate the two experiments 1 and 2 (yielding the relative likelihoods $p_1/q$ and $p_2/q$ respectively) graded with the same $c_{\rm novel}$ into an effective single test with the same $c_{\rm novel}$ and likelihood $(p_1/q)(p_2/q)$. This is the ideal situation, as there are cases where repeating an experiment may wildly increase the evidence of systemic uncertainty or demonstrate uncontrolled variability or other kinds of problems. When this occurs, this means that the assumption that there is no surprise, no novelty, in repeating the experiment is incorrect. Then, the two experiments should be treated so as to contribute two multipliers $F$'s, because they reveal different kinds of uncertainty that can be generated by ensembles of experiments.

One experimental test corresponds to a entire loop $1-4$ transforming a given $V_{\rm prior}$ to a $V_{\rm posterior}$ according to (\ref{mgler}). This $V_{\rm posterior}$ becomes the new $V_{\rm prior}$ for the next test, which will transform it into another $V_{\rm posterior}$ and so on, according to the following iteration process:
\begin{equation}
V_{\rm prior}^{(1)} \to
V_{\rm posterior}^{(1)} = V_{\rm prior}^{(2)} \to
V_{\rm posterior}^{(2)} = V_{\rm prior}^{(3)} \to \cdots \to
V_{\rm posterior}^{(n)}~.
\label{mgmbmel}
\end{equation}
After $n$ validation loops, we have a posterior trust in the model given by
\footnote{
This sequence is reminiscent of a branching process: most of the time, after the first or second validation loop, the model will be rejected if $V_{\rm posterior}^{(n)}$ becomes much smaller than $V_{\rm prior}^{(1)}$. The occurrence of a long series of validation tests is specific to those rare models/codes that happen to survive. We conjecture that the nature of models and their tests make the probability of survival up to level $n$ a power law decaying as a function of validation generation number $n$:
${\rm Pr}\left[{ V_{\rm posterior}^{(n)} \geq V_{\rm prior}^{(1)} }\right] \sim {1/n^{\tau}}$, for large $n$.  
The exponent $\tau = 3/2$ in mean-field branching processes \cite{Har99}; being an ensemble average over random test outcomes, we expect this to be only an upper bound for actual validation processes.  The four illustrative examples provided further on, augmented with a fifth one described in Ref. \cite{Sor07}, yield $\tau \approx 0.85$ for $3 \le n \le 7$ with just one outlier.  Although the sample of models is tiny, this illustrates our point.
}
\begin{equation}
\frac{ V_{\rm posterior}^{(n)} }{ V_{\rm prior}^{(1)} } =
F\left[ p^{(1)}(M | y^{(1)}_{\rm obs}), q^{(1)}; c^{(1)}_{\rm novel} \right]
~\cdots~
F\left[ p^{(n)}(M | y^{(n)}_{\rm obs}), q^{(n)}; c^{(n)}_{\rm novel} \right]~,
\label{mglerdffd}
\end{equation}
where the product is time-ordered since the sequence of values for $c^{(j)}_{\rm novel}$ depend on preceding tests. Validation can be said to be asymptotically satisfied when the number of steps $n$ and the final value $V_{\rm posterior}^{(n)}$ are sufficiently high. How high is high enough is subjective and may depend on both the application and programmatic constraints. The concrete examples discussed below offer some insight on this issue. This construction makes clear that there is no absolute validation, only a process of corroborating or disproving steps competing in a global valuation of the model under scrutiny. The product (\ref{mglerdffd}) expresses the assumption that successive observations give independent multipliers. This assumption keeps the procedure simple because determining the dependence between different tests with respect to validation would be highly undetermined. We propose that it is more convenient to measure the dependence through the single parameter $c^{(j)}_{\rm novel}$ quantifying the novelty of the $j$th test with respect to those preceding it. In full generality, each new $F$ multiplier should be a function of all previous tests.

The loop $1-4$ together with expression (\ref{mgler}) are offered as an attempt to quantify the progression of the validation process. Eventually, when one has performed several approximately independent tests exploring different features of the model and of the validation process, $V_{\rm posterior}$ has grown to a level at which most experts will be satisfied and will believe in the validity of (i.e., be inclined to trust) the model. This formulation has the advantage of viewing the validation process as a convergence or divergence built on a succession of steps, mimicking the construction of a theory of reality.
\footnote{ 
It is conceivable that a new and radically different
observation/experiment may arise and challenge the built-up trust in a model;
such a scenario exemplifies how any notion of validation ``convergence'' is
inherently local.
}
Expression (\ref{mglerdffd}) embodies the progressive build-up of trust in a model or theory. This formulation provides a formal setting for discussing the difficulties that underlay the so-called impossibilities \cite{Ore94a,Ryk94} in validating a given model. Here, these difficulties are not only partitioned but quantified:
\begin{itemize}
\item
in the definition of ``new'' non-redundant experiments (parameter $c_{\rm novel}$),
\item
in choosing the metrics and the corresponding statistical tests quantifying the comparison between the model and the measurements of this experiment (leading to the likelihood ratio $p/q$), and
\item
in iterating the procedure so that the product of the gain/loss factors $F[\cdots]$ obtained after each test eventually leads to a clear-cut conclusion after several tests.
\end{itemize}
This formulation makes clear why and how one is never fully convinced that validation has been obtained: it is a matter of degree, of confidence level, of decision making, as in statistical testing. But this formulation helps in quantifying what new confidence (or distrust) is gained in a given model. It emphasizes that validation is an ongoing process, similar to the never-ending construction of a theory of reality.

The general formulation proposed here in terms of iterated validation loops is intimately linked with decision theory based on limited knowledge: the decision to ``go ahead'' and use the model is fundamentally a decision problem based on the accumulated confidence embodied in $V_{\rm posterior}$. The ``go/no-go'' decision must take into account conflicting requirements and compromise between different objectives. Decision theory was created by the statistician Abraham Wald in the late forties \cite{Wal50}, but is based ultimately on game theory \cite{Neu44,Kot93}. Wald used the term {\it loss function}, which is the standard terminology used in mathematical statistics. In mathematical economics, the opposite of the loss (or cost) function gives the concept of the {\it utility function}, which quantifies (in a specific functional form) what is considered important and robust in the fit of the model to the data. We use $V_{\rm posterior}$ in an even more general sense than ``utility,'' as a decision and information-based valuation that supports risk-informed decision-making based on ``satisficing''
\footnote{
In economics, {\it satisficing} is a behavior that attempts to achieve at least some minimum level of a particular variable, but that does not strive to achieve its maximum possible value. The verb ``to satisfice'' was coined by Herbert A. Simon in his theory of bounded rationality \cite{Sim55,Sim79}.
}
(see the concrete examples discussed below).

It may be tempting to interpret the above formulation of the validation problem in terms of Bayes' theorem
\begin{equation}
p_{\rm posterior}(M | {\rm Data}) = 
\frac{ p_{\rm prior}(M) \times \Pr({\rm Data}|M) }{ \Pr({\rm Data}) }
\label{gnnfel}
\end{equation}
where $\Pr({\rm Data}|M)$ is the likelihood of the data given  the model $M$, and $\Pr({\rm Data})$ is the unconditional likelihood of the data.   However, we can not make immediate sense of $\Pr({\rm Data})$.  Only when a second model $M'$ is introduced can we actually calculate
\begin{equation}
{\rm Pr}({\rm Data}) = p_{\rm prior}(M) ~{\rm Pr}({\rm Data}|M)  +
p_{\rm prior}(M') ~{\rm Pr}({\rm Data}|M') ~.
\end{equation}
In other words, Bayes' formulation requires that we set a  model/hypothesis in opposition to another or other ones, while we examine here the case of a single hypothesis in isolation.

We therefore stress that one should resist the urge to equate our $V_{\rm prior}$ and $V_{\rm posterior}$ with $p_{\rm prior}$ and $p_{\rm posterior}$ because they are {\it not} probabilities. It is not possible to assign a probability to an experiment in an absolute way and thus Bayes' theorem is mute on the validation problem as we have chosen to formulate it. Rather, we propose that the problem of validation is fundamentally a problem of decision theory: at what stage is one willing to bet that the code will work for its intended use? At what stage, are you ready to risk your reputation, your job, the lives of others, your own life on the fact that the model/code will predict correctly the crucial aspect of the real-life test? One must therefore incorporate ingredients of decision theory, and not only fully objective probabilities. Coming from a Bayesian perspective, $p_{\rm prior}$ and $p_{\rm posterior}$ could then be called the potential {\it value} or {\it trust} in the model/code or, as we prefer, to move closer to the application of decision theory in economics, the {\it utility} of the model/code \cite{Neu44}.

To summarize the discussion so far, expression (\ref{mgler}) may be reminiscent of a Bayesian analysis, however, it does not manipulate probabilities. (Instead, they appear as {\it independent} variables, viz., $p(M | y_{\rm obs})$ and $q$.) In the Bayesian methodology of validation \cite{Zha03,Mah05}, only comparison between models can be performed due to the need to remove the unknown probability of the data in Bayes' formula. In contrast, our approach provides a value for each single model independently of the others. In addition, it emphasizes the importance of quantifying the novelty of each test and takes a more general view on how to use the information provided from the goodness-of-fit. The valuation (\ref{mgler}) of a model uses probabilities as partial inputs, not as the qualifying criteria for model validation. This does not mean, however, that there are not uncertainties in these quantities or in the terms $F$, $q$ or $c_{\rm novel}$ and that aleatory and epistemic uncertainties
\footnote{
For an in-depth discussion on aleatory versus systemic (a.k.a. epistemic) uncertainties, see for example {\it Review of Recommendations for Probabilistic Seismic Hazard Analysis: Guidance on Uncertainty and Use of Experts} \cite{NRC97}, available at http://www.nap.edu/catalog/5487.html.
}
are ignored, as discussed below.

\section{Desirable Properties of the Multiplier of the Validation Step}

The multiplier $F\left[ p(M | y_{\rm obs}), q; c_{\rm novel} \right]$ should have the following properties:
\begin{enumerate}
\item
If the statistical test(s) performed on the given observations is (are) passed at the reference level $q$, then the posterior potential value is larger than the prior potential value: $F>1$ (resp. $F \leq 1$) for $p>q$ (resp. $p \leq q$), which can be written succinctly as $\ln F / \ln (p/q) > 0$.
\item
The larger the statistical significance of the passed test, the larger the posterior value. Hence
\begin{equation*}
\frac{ \partial F }{ \partial p } > 0~,
\end{equation*}
for a given $q$. There could be a saturation of the growth of $F$ for large $p/q$, which can be either that $F < \infty$ as $p/q \to \infty$ or of the form of a concavity requirement
\begin{equation*}
\frac{ \partial^2 F }{ \partial p^2 } < 0
\end{equation*}
for large $p/q$: obtaining a quality of fit beyond a certain level should not be attempted.
\item
The larger the statistical level at which the test(s) performed on the given observations is (are) passed, the larger the impact of a ``novel'' experiment on the multiplier enhancing the prior into the posterior potential value of the model: $\partial F / \partial c_{\rm novel} > 0$ (resp. $\leq 0$), for $p>q$ (resp. $p \leq q$).
\end{enumerate}

\begin{figure}
\begin{center}
\includegraphics[width=12cm]{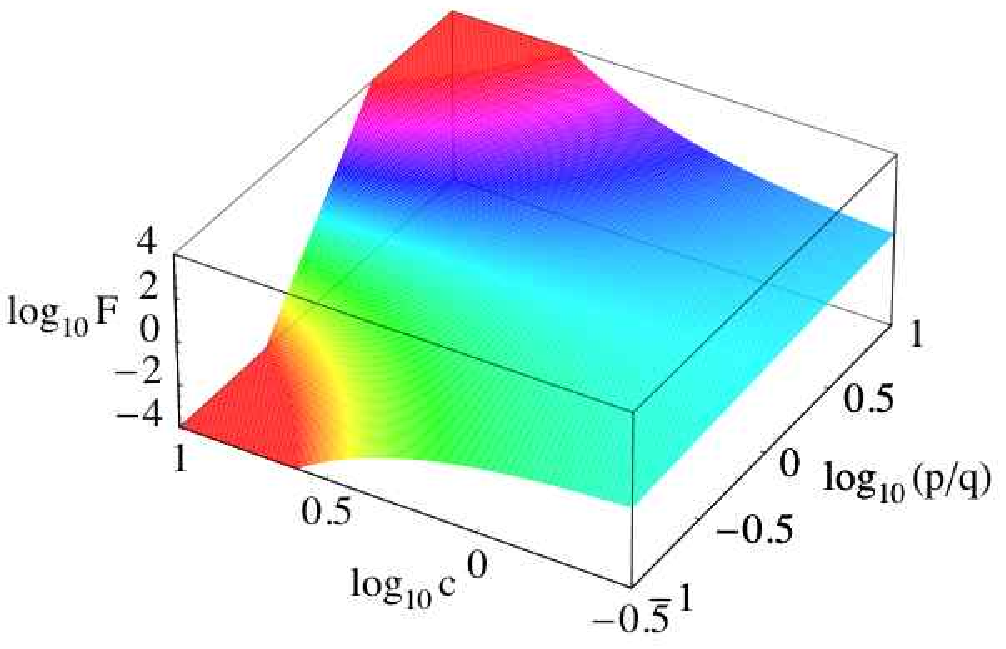}
\includegraphics[width=12cm]{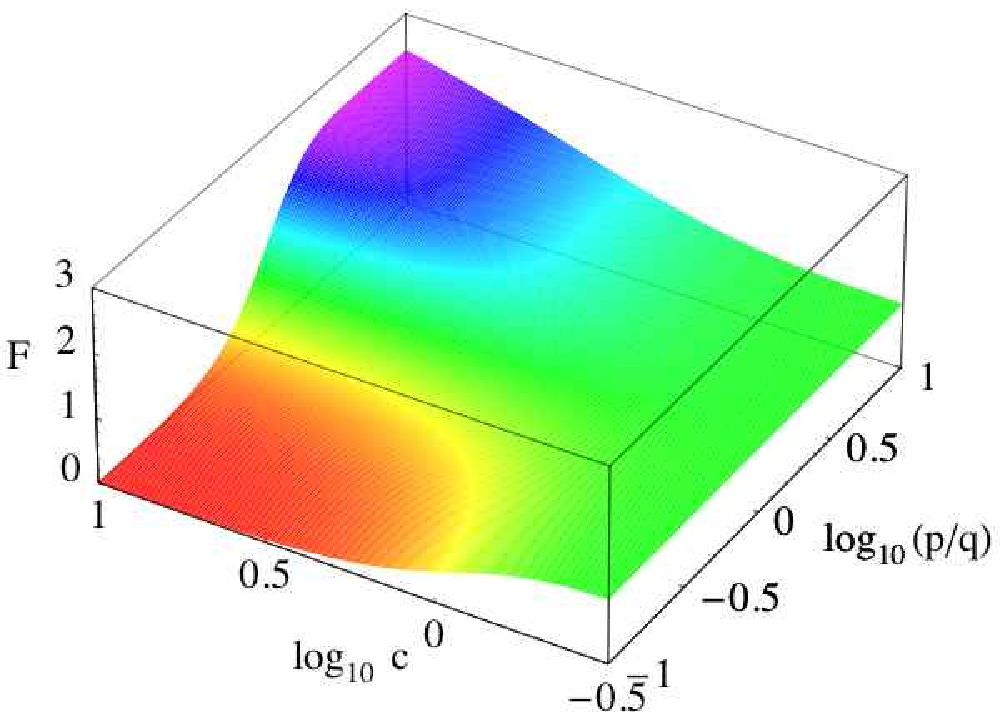}
\caption{The multipliers defined by (\ref{mfmsl}) and (\ref{mgler222}) are plotted as functions of $p/q$ and $c_{\rm novel}$ in the upper and lower panels respectively.  Note the vertical log scale used for the multiplier (\ref{mfmsl}) in the top panel.}
\label{Fmult}
\end{center}
\end{figure}

A very simple multiplier that obeys this these properties (not including the saturation of the growth of $F$) is given by
\begin{equation}
F\left[ p(M | y_{\rm obs}), q; c_{\rm novel} \right] =
\left( \frac{p}{q} \right)^{c_{\rm novel}}~,
\label{mfmsl}
\end{equation}
and is illustrated in the upper panel of Fig.~\ref{Fmult} as a function of $p/q$ and $c_{\rm novel}$. This form provides an intuitive interpretation of the meaning of the experiment impact parameter $c_{\rm novel}$. A non-committal evaluation of the novelty of a test would be $c_{\rm novel} = 1$, thus $F = p/q$ and the chain (\ref{mglerdffd}) reduces to a product of normalized likelihoods, as in standard statistical tests. A value $c_{\rm novel} > 1$ (resp. $< 1$) for a given experiment describes a nonlinearly rapid (resp. slow) updating of our trust $V$ as a function of the grade $p/q$ of the model with respect to the observations. In particular, a large value of $c_{\rm novel}$ corresponds to the case of ``critical'' tests.
\footnote{A momentous example is the Michelson-Morley experiment for the Theory of Special Relativity. For the Theory of General Relativity, it was the observation during the famous 1919 solar eclipse of the bending of light rays from distant stars by the Sun's mass and the elegant explanation of the anomalous precession of the perihelion of Mercury's orbit.
}
Note that the parameterization of $c_{\rm novel}$ in (\ref{mfmsl}) should account for the decreased novelty noted above occurring when the same experiment is repeated two or more times. The value of $c_{\rm novel}$ should be reduced for each repetition of the same test; moreover, the value of $c_{\rm novel}$ should approach unity as the number of repetitions increases.

An alternative multiplier,
\begin{equation}
F\left[ p(M | y_{\rm obs}), q; c_{\rm novel} \right] =
\left[ \frac{ \tanh \left( \frac{p}{q} + \frac{1}{c_{\rm novel}} \right) }
            { \tanh \left( 1           + \frac{1}{c_{\rm novel}} \right) } 
\right]^4~,
\label{mgler222}
\end{equation}
is plotted in the lower panel of Fig.~\ref{Fmult} as a function of $p/q$ and $c_{\rm novel}$. It emphasizes that $F$ saturates as a function of $p/q$ and $c_{\rm novel}$ as either one or both of them grow large. A completely new experiment corresponds to $c_{\rm novel} \to \infty$ so that $1/c_{\rm novel}=0$ and thus $F$ tends to $[\tanh(p/q)/\tanh(1)]^4$, i.e., $V_{\rm posterior}/V_{\rm prior}$ is only determined by the quality of the ``fit'' of the data by the model quantified by $p/q$.  A finite $c_{\rm novel}$ thus implies that one already takes a restrained view on the usefulness of the experiment since one limits the amplitude of the ${\rm gain}=V_{\rm posterior}/V_{\rm prior}$, whatever the quality of the fit of the data by the model. The exponent $4$ in (\ref{mgler222}) has been chosen so that the maximum confidence gain $F$ is equal to $\tanh(1)^{-4} \approx 3$ in the best possible situation of a completely new experiment ($c_{\rm novel} = \infty$) and perfect fit ($p/q \to \infty$). In contrast, the multiplier $F$ can be arbitrarily small as $p/q \to 0$ even if the novelty of the test is high ($c_{\rm novel} \to \infty$). For a finite novelty $c_{\rm novel}$, a test that fails the model miserably ($p/q \approx 0$) does not necessarily reject the model completely: unlike the expression in (\ref{mfmsl}), $F$ remains greater than zero.  Indeed, if the novelty $c_{\rm novel}$ is small, the worst-case multiplier (attained for $p/q = 0$) is 
$\left[
\tanh \left( 1/c_{\rm novel} \right) /
\tanh \left( 1+1/c_{\rm novel} \right)
\right]^4 \approx 1 - 6.9\,{\rm e}^{-2/c_{\rm novel}}$,
which is only slightly less than unity if $c_{\rm novel} \ll 1$. In short, this formulation does not heavily weight unimportant tests,  as seems intuitively appropriate.

In the framework of decision theory, expression (\ref{mgler}) with one of the specific expressions in (\ref{mfmsl}) or (\ref{mgler222}) provides a parametric form for the utility or decision ``function'' of the decision maker. It is clear that many other forms of the utility function can be used, however, with the constraint of keeping the salient features of expression (\ref{mgler}) with (\ref{mfmsl}) or (\ref{mgler222}), in terms of the impact of a new test given past tests, and the quality of the comparison between the model predictions and the data. This indetermination is helpful since it mirrors the inherent variability of the validation landscape.  For instance, what comprises adequate validation for phenomena at one (e.g., macro-)scale may prove inadequate for related phenomena at another (e.g., micro-)scale.

Finally, we remark that the proposed form for the multiplier (\ref{mgler222}) contains an important asymmetry between gains and losses: the failure to a single test with strong novelty and significance
\footnote{
See, e.g., the impact of localized seismicity on faults in the case of the Olami-Feder-Christensen model discussed below, or that of the ``leverage'' effect in quantitative finance for the Multifractal Random Walk model described and evaluated in Ref.~\cite{Sor07}.
}
cannot be compensated by the success of all the other tests combined. In other words, a single test is enough to reject a model. This encapsulates the common lore that reputation gain is a slow process requiring constancy and tenacity, while its loss can occur suddenly with one single failure and is difficult to re-establish. We believe that the same applies to the build-up of trust in and, thus, validation of a model.

\section{Practical Guidelines for Determining $p/q$ and $c_{\rm novel}$}

These two crucial elements of a validation step are conditioned by four basic problems, over which one can exert at least partial control. In particular, they address the two sources of uncertainty: ``reducible'' or epistemic (i.e., due to lack of knowledge) and ``irreducible'' or aleatory (i.e., due to variability inherent in the phenomenon under consideration). In a nutshell, as becomes clearer below, the comparison between $p$ and $q$ is more concerned with the aleatory uncertainty while $c_{\rm novel}$ deals in part with the epistemic uncertainty. In the following, as in the two examples (\ref{mfmsl}) and (\ref{mgler222}), we consider that $p$ and $q$ enter only in the form of their ratio $p/q$. This should not be generally the case but, given the many uncertainties, this restriction simplifies the analysis by removing one degree of freedom.

\begin{enumerate}

\item {\it How to model?}
This addresses model construction and involves the structure of the elementary contributions, their hierarchical organization, and requires dealing with uncertainties and fuzziness. This concerns the epistemic uncertainty.

\item {\it What to measure?}
This relates to the nature of $c_{\rm novel}$: ideally, one should target adaptively the observations to ``sensitive'' parts of the system and the model (as, e.g., Palmer et al. \cite{Pal98} did for atmospheric dynamics). Targeting observations could be directed by the desire to access the most ``relevant'' information as well as to get information that is the most reliable, i.e., which is contaminated by the smallest errors. This is also the stance of Oberkampf and Trucano \cite{Obe02a}: {\it ``A validation experiment is conducted for the primary purpose of determining the validity, or predictive accuracy, of a computational modeling and simulation capability. In other words, a validation experiment is designed, executed, and analyzed for the purpose of quantitatively determining the ability of a mathematical model and its embodiment in a computer code to simulate a well-characterized physical process.''} In practice, we view $c_{\rm novel}$ as an estimate of the importance of the new observation and the degree of ``surprise'' it brings to the validation step. Being the cornerstone of our formal approach to validation, we eventually want to see its determination grounded in sensitivity and/or PIRT analysis (cf. section~\ref{s:cal+DA}).  The epistemic uncertainty alluded to above is partially addressed in the choice of the empirical data and its rating with $c_{\rm novel}$ (see the examples of application discussed below).

\item {\it How to measure?}
For given measurements or experiments, the problem is to find the ``optimal'' metric or cost function (involved in the quality-of-fit measure $p$) for the intended use of the model. The notion of optimality needs to be defined. It could capture a compromise between fitting best the important features of the data (what is ``important'' may be decided on the basis of previous studies and understanding or other processes, or programmatic concerns), and minimizing the extraction of spurious information from noise. This requires one to have a precise idea of the statistical properties of the noise. If such knowledge is not available, the cost function should be chosen accordingly. The choice of the cost function involves the choice of how to look at the data. For instance, one may want to expand the measurements at multiple scales using wavelet decompositions and compare the prediction and observations scale by scale, or in terms of multifractal spectra of the physical fields estimated from these wavelet decompositions \cite{Muz94} or from other methods. The general idea here is that, given complex observation fields, it is appropriate to unfold the data on a variety of ``metrics,'' which can then be used in the comparison between observations and model predictions. The question is then: How well is the model able to reproduce the salient multi-scale and multifractal properties derived from the observations? The physics of turbulent fields and of complex systems have offered many such new tools with which to unfold complex fields according to different statistics. Each of these statistics offers a metric to compare observations with model predictions and is associated with a cost function focusing on a particular feature of the process. Since these metrics are derived from the understanding that turbulent fields can be analyzed using these metrics that reveal strong constraints in their organization, these metrics can justifiably be called ``physics-based.'' In practice, $p$, and eventually $p/q$, has to be inferred as an estimate of the degree of matching between the model output and the observation. This can be done following the concept of fuzzy logic in which one replaces the yes/no pass test by a more gradual quantification of matching \cite{Bel70,Zad92}. We thus concur with Oberkampf and Barone \cite{Obe05}, while our general methodology goes beyond. Note that this discussion relates primarily to the aleatory uncertainty.

\item {\it How to interpret the results?}
This question relates to defining the test and the reference probability level $q$ that any other model (than the one under scrutiny) can explain the data. The interpretation of the results should aim at detecting the ``dimensions'' that are missing, misrepresented or erroneous in the model (systemic/epistemic uncertainty). What tests can be used to betray the existence of hidden degrees of freedom and/or dimensions? This is the hardest problem. It can sometimes possess an elegant solution when a given model is embedded in a more general one. Then, the limitation of the more restricted model becomes clear from the vantage of the more general model.

\end{enumerate}

\noindent
We refer to the Appendix for further thoughts on these four basic steps in model construction and validation in a broader context than our present formulation.

We now illustrate our algorithmic approach to model validation using the historical development of quantum mechanics and three examples based on the authors' research activities. In these crude but revealing examples, we will use the form (\ref{mgler222}) and consider three finite values:
$c_{\rm novel}=1$ (marginally useful new test),
$c_{\rm novel}=10$ (substantially new test), and
$c_{\rm novel}=100$ (important new test).
When a likelihood test is not available, we propose to use three possible marks:
$p/q=0.1$ (poor fit),
$p/q=1$ (marginally good fit), and
$p/q=10$ (good fit).
Extreme values ($c_{\rm novel}$ or $p/q$ are 0 or $\infty$) have already been discussed.  Due to limited experience with this approach, we propose these ad hoc values in the following examples of its application.

\section{Illustration with the Development of Quantum Mechanics}

Quantum mechanics (QM) offer a vivid incarnation of how a model can turn progressively into a theory held ``true'' by almost all physicists. Since its birth, QM has been tested again and again because it presents a view of ``reality'' that is shockingly different from the classical view experienced at the macroscopic scale. QM prescriptions and predictions often go against (classically-trained) intuition. Nevertheless, we can state that, by a long and thorough process of confirmed predictions of QM in experiments, fueled by the imaginative set-up of paradoxes, QM has been validated as a correct description of nature. It is fair to say that the overwhelming majority of physicists have developed a strong trust in the validity of QM. That is, if someone comes up with a new test based on a new paradox, most physicists would bet that QM will come up with the right answer with a very high probability. It is thus by the on-going testing and the compatibility of the prediction of QM with the observations that QM has been validated. As a consequence, one can use it with strong confidence to make predictions in novel directions. This is ideally the situation one would like to attain for the problem of validation of all models, those discussed in the following section in particular. We now give a very partial list of selected tests that established the trust of physicists in QM.

\begin{enumerate}

\item Pauli's exclusion principle states that no two identical fermions (particles with non-integer values of spin) may occupy the same quantum state simultaneously \cite{Mas05}. It is one of the most important principles in quantum physics, primarily because the three types of particle from which ordinary matter is made, electrons, protons, and neutrons, are all subject to it. With $c_{\rm novel} = 100$ and perfect agreement in numerous experiments ($p/q = \infty$), this leads to $F^{(1)}=2.9$.

\item The EPR paradox \cite{Ein35} was a thought experiment designed to prove that quantum mechanics was hopelessly flawed: according to QM, a measurement performed on one part of a quantum system can have an instantaneous effect on the result of a measurement performed on another part, regardless of the distance separating the two parts. Bell's theorem \cite{Bel64} showed that quantum mechanics predicted stronger statistical correlations between entangled particles than the so-called local realistic theory with hidden variables. The importance of this prediction requires $c_{\rm novel} = 100$ at a minimum. The QM prediction turned out to be correct, winning over the hidden-variables theories \cite{Asp80,Rar90} ($p/q = \infty$), leading again to $F^{(2)}=2.9$.

\item The Aharonov-Bohm effect predicts that a magnetic field can influence an electron that, strictly speaking, is located completely beyond the field's range, again an impossibility according to non-quantum theories ($c_{\rm novel} = 100$). The Aharonov-Bohm oscillations were observed in ordinary (i.e., not superconducting) metallic rings, showing that electrons can maintain quantum mechanical phase coherence in ordinary materials \cite{Web85,Sch86}. This yields $p/q = \infty$ and thus $F^{(3)}=2.9$ yet again.

\item The Josephson effect provides a macroscopic incarnation of quantum effects in which two superconductors are predicted to preserve their long-range order across an insulating barrier, for instance, leading to rapid alternating currents when a steady voltage is applied across the superconductors. The novelty of this effect again warrants $c_{\rm novel} = 100$ and the numerous verifications and applications (for instance in SQUIDs, {S}uperconducting {QU}antum {I}nterference {D}evices) argues for $p/q = \infty$ and thus $F^{(4)}=2.9$, as usual.

\item The prediction of possible collapse of a gas of atoms at low temperature into a single quantum state is known as Bose-Einstein condensation, again so much against classical intuition ($c_{\rm novel} = 100$). Atoms are indeed bosons (particles with integer values of spin), which are {\it not} subjected to the Pauli exclusion principle evoked in the above test \#1 of QM.  The first such Bose-Einstein condensate was produced using a gas of rubidium atoms cooled to $1.7 \cdot 10^{-7}$ K \cite{And95} ($p/q = \infty$), leading once more to $F^{(4)}=2.9$.

\item There have been several attempts to develop a paradox-free nonlinear QM theory, in the hope of eliminating Schr\"odinger's cat paradox, among other embarrassments. The nonlinear QM predictions diverge from those of orthodox quantum physics, albeit subtly. For instance, if a neutron impinges on two slits, an interference pattern appears, which should, however, disappear if the measurement is made far enough away ($c_{\rm novel} = 100$). Experiment tests of the neutron prediction rejected the nonlinear version in favor of the standard QM \cite{Gah81} ($p/q = \infty$), leading to $F^{(6)}=2.9$.

\item In addition, measurements at the National Institute of Standards and Technology (NIST) in Boulder, CO, on frequency standards have been shown to set limits of order $10^{-21}$ on the fraction of the energy of the rf transition in $^9$Be ions that could be due to nonlinear corrections to quantum mechanics \cite{Wei89}. We assign $c_{\rm novel} = 10$, with $p/q = 10$), to this result, leading to $F^{(7)}=2.4$.  Although less than $F^{(1-6)}$, this is still an impressive score.

\end{enumerate}

\noindent 
Combining the multipliers according to (\ref{mglerdffd}) leads to $V_{\rm posterior}^{(8)} / V_{\rm prior}^{(1)} \simeq 1400$, which is of course only a lower limit given the many other validation tests not mentioned here.

Tests of QM are ongoing \cite{Leg02}. But given the presumably huge amount of trust physicists have in QM which we tried to quantify, why do physicists still feel the need to put QM to the ``validation'' test? This raises the question whether we can ever establish a sense of sufficiency for validation. Our position is that this reflects a quixotic quest for absolute truth---and also a taste for surprises---that most scientists can relate to. Perhaps, by continuing to test QM, a new insight or an anomaly will be uncovered which may help progress in the understanding of reality.

\section{Three Examples Drawn from the Authors' Research Interests}

\subsection{The Olami-Feder-Christensen (OFC) Sand-Pile Model of Earthquakes}

\begin{figure}
\begin{center}
\includegraphics[width=12cm]{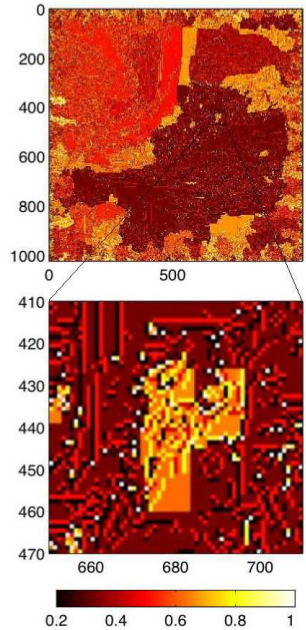}
\caption{Map of the ``stress'' field generated by the OFC model immediately after a large avalanche (main shock) at two magnifications. The upper panel shows the whole grid of size 1024 and the lower plot represents a subset of the grid delineated by the square in the upper plot. Adapted from Ref.~\cite{Hel04}.}
\label{Figure_OFC}
\end{center}
\end{figure}

This is perhaps the simplest sand-pile model of self-organized criticality, which exhibits a phenomenology resembling real seismicity \cite{Ola92}. Figure~\ref{Figure_OFC} shows a ``stress'' map generated by the OFC model immediately after a large avalanche (main shock) at two magnifications, to illustrate the rich organization of almost synchronized regions \cite{Dro02}. To validate the OFC model, we examine the properties and prediction of the model that can be compared with real seismicity, together with our assessment of their $c_{\rm novel}$ and quality-of-fit. We are careful to state these properties in an ordered way, as specified in the above sequences (\ref{mgmbmel})--(\ref{mglerdffd}).

\begin{enumerate}

\item The statistical physics community recognized the discovery of the OFC model as an important step in the development of a theory of earthquakes: without a conservation law (which was thought before to be an essential condition), it nevertheless exhibits a power law distribution of avalanche sizes resembling the Gutenberg-Richter law \cite{Ola92}. On the other hand, many other models with different mechanisms can explain observed power law distributions \cite{Sor04}. We thus attribute only $c_{\rm novel} = 10$ to this evidence. Because the power law distribution obtained by the model is of excellent quality for a certain parameter value ($\alpha \approx 0.2$), we formally take $p/q = \infty$ (perfect fit). Expression (\ref{mgler222}) then gives $F^{(1)}=2.4$.

\item Prediction of the OFC model concerning foreshocks and aftershocks, and their exponents for the inverse and direct Omori laws. These predictions are twofold \cite{Hel04}: (i) the finding of foreshocks and aftershocks with similar qualitative properties, and (ii) their inverse and direct Omori rates. The first aspect, deserves a large $c_{\rm novel}=100$ as the observation of foreshocks and aftershocks came as a rather big surprise in such sand-pile models \cite{Her02}. The clustering in time and space of the foreshocks and aftershocks are qualitatively similar to real seismicity \cite{Hel04}, which warrants $p/q=10$, and thus $F^{(2a)}=2.9$. The second aspect is secondary compared with the first one ($c_{\rm novel}=1$). Since the exponents are only qualitatively reproduced (but with no formal likelihood test available), we therefore take $p/q=0.1$. This leads to $F^{(2b)}=0.47$.

\item Scaling of the number of aftershocks with the main shock size (productivity law) \cite{Hel04}: $c_{\rm novel}=10$ as this observation is rather new but not completely independent of the Omori law. The fit is good so we grant a grade $p/q=10$ leading to $F^{(3)}=2.4$. \item Power law increase of the number of foreshocks with the main shock size \cite{Hel04}: this is not observed in real seismicity, probably because this property is absent or perhaps due to a lack of quality data. This test is therefore not very selective ($c_{\rm novel}=1$) and the large uncertainties suggest a grade $p/q=1$ (to reflect the different viewpoints on the absence of effect in real data) leading to $F^{(4)}=1$ (neutral test).

\item Most aftershocks are found to nucleate at ``asperities'' located on the main shock rupture plane or on the boundary of the avalanche, in agreement with observations \cite{Hel04}: $c_{\rm novel}=10$ and $p/q=10$ leading to $F^{(5)}=2.4$.

\item Earthquakes cluster on spatially localized geometrical structures known as faults. This property is arguably central to the physics of seismicity ($c_{\rm novel}=100$), but absolutely not reproduced by the OFC model ($p/q=0.1$). This leads to $F^{(6)}= 4 \cdot 10^{-4}$.

\end{enumerate}

\noindent 
Combining the multipliers according to (\ref{mglerdffd}) up to test \#5 leads to $V_{\rm posterior}^{(6)} / V_{\rm prior}^{(1)} =  18.8$, suggesting that the OFC model is validated as a useful model of the statistical properties of seismic catalogs, at least with respect to the properties which have been examined in these first five tests. Adding the crucial last test strongly fails the model since $V_{\rm posterior}^{(7)} / V_{\rm prior}^{(1)} = 7.5\,10^{-3}$. The model can not be used as a realistic predictor of seismicity. The results of our quantitative validation process indicate that it can nevertheless be useful to illustrate certain statistical properties and to help formulate new questions and hypotheses.

\subsection{An Anomalous Diffusion Model for Solar Radiation in Cloudy Atmospheres}

\begin{figure}
\begin{center}
\includegraphics[width=8cm]{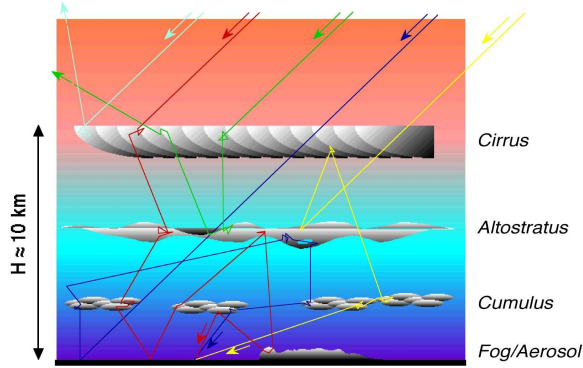}
\caption{
Schematic representation of the anomalous diffusion model of solar photon transport at non-absorbing wavelengths in the cloudy atmosphere. In this model, solar beams follow convoluted L\'evy walks, which are characterized by frequent small steps (inside clouds) and occasional large jumps (between clouds or between clouds and the surface). The partition between small and large jumps is controlled by the L\'evy index $\alpha$ (the PDF of the jump sizes $\ell$ has a tail decaying as a power law $\sim 1/\ell^{1+\alpha}$). Reproduced from Ref.~\cite{Dav06}.
}
\label{anomalous_diffusion}
\end{center}
\end{figure}

To improve our modeling skill for climate dynamics, it is essential to reduce the significant uncertainty associated with clouds. In particular, estimation of the radiation budget in the presence of clouds needs improvement since current operational models for the most part ignore all variability below the scale of the climate model's grid ($\sim$100~km).  A considerable effort has therefore been expended to derive more realistic mean-field radiative transfer models \cite{Bar05}, mostly by considering only the one-point variability of clouds, that is, irrespective of their actual structure as captured by 2-point (or higher) correlation statistics.  However, it has been widely recognized that the Earth's cloudiness is fractal over a wide range of scales \cite{Lov82}. This is the motivation for modeling the paths of solar photons at non-absorbing wavelengths in the cloudy atmosphere as L\'evy walks \cite{Sor04}, which are characterized by frequent small steps (inside clouds) and occasional large jumps (typically between clouds) as represented schematically in Fig.~\ref{anomalous_diffusion}. These (on-average downward) paths start at the top of the highest clouds and end in escape to space or in absorption at the surface, respectively, cooling and warming the climate system. In contrast with most other mean-field models for solar radiative transfer, this diffusion model with anomalous scaling can be subjected to a battery of observational tests.

\begin{enumerate}

\item The original goal of this phenomenological model, which accounts for the clustering of cloud water droplets into broken and/or multi-layered cloudiness, was to predict the increase in steady-state flux transmitted to the surface compared to what would filter through a fixed amount of condensed water in a single unbroken cloud layer \cite{Dav97}.  This property is common to all mean-field photon transport models that do anything at all about unresolved variability \cite{Bar05}. Thus, we assign only $c_{\rm novel}=1$ to this test and, given that all models in this class are successful, we have to take $p/q=1$, hence $F^{(1)}=1$.  The outcome of this first test is neutral.

\item The first real test for this model occurred in the late 1990s, when it became possible to accurately estimate the mean total path cumulated by solar radiation that reaches the surface.  This breakthrough was enabled by access to spectroscopy at medium (high) resolution  of oxygen bands (lines) \cite{Pfe99,Min01}.  There was already remote sensing technology to infer simultaneously cloud optical depth, which is column-integrated water in g (or cm$^3$) per cm$^2$ multiplied by the average cross-section for scattering or absorption in cm$^2$ per g (or cm$^3$).  The observed trends between mean path and optical depth were explained only by the new model in spite of relatively large instrumental error bars.  So we assign $c_{\rm novel}=100$ to this highly discriminating test and $p/q=10$ (even though other models were generally not in a position to compete), hence $F^{(2)}=2.9$.

\item Another test was proposed using time-dependent photon transport with a source near the surface (cloud-to-ground lightning) and a detector in space (aboard the US DOE FORT\'{E} satellite) \cite{Dav00}.  The quantity of interest is the observed delay of the light pulse (due to multiple scattering in the cloud system) with respect to the radio-frequency pulse (which travels in a straight line).  There was no simultaneous estimate of cloud optical depth, so assumptions had to be made (informed by the fact that storm clouds are at once thick and dense).  Because of this lack of an independent measurement, we assign only $c_{\rm novel}=10$ to the observation and $p/q=1$ to the model performance.  Indeed, this test is arguably only about the finite horizontal extent of the rain clouds resulting from deep convection: one can exclude only most simplistic cloud models based on uniform plane-parallel slabs.  So, again we obtain $F^{(3)}=1$ for an interesting but presently neutral test that needs refinement.

\item Min et al. \cite{Min04} developed an oxygen-line spectrometer with sufficient resolution to estimate not just the {\it mean} path but also its {\it root-mean-square} (RMS) value.  They found the prediction by Davis and Marshak \cite{Dav02} for normal diffusion to be an extreme (envelop) case for the empirical scatter plot of mean vs. RMS path, and this is indicative that the anomalous diffusion model will cover the bulk of the data.  Because of some overlap with item \#2, we assign $c_{\rm novel}=10$ to the test and $p/q=10$ for the model performance since the anomalous diffusion model had not yet made a prediction for the RMS path (although we note that other models have yet to make one for the mean path). We therefore receive $F^{(4)}=2.4$.

\item Using similar data but a different normalization than Min et al., more amenable to model testing, Scholl et al. \cite{Sch06} observed that the RMS-to-mean ratio for solar photon path is essentially constant whether the cloud structure (according to mm-wave radar profiles) is complex or not (respectively, diffusion is normal or anomalous).  This is a remarkable empirical finding to which we assign $c_{\rm novel}=100$.  The new mean- and RMS-path data was explained by Scholl et al. by creating an ad hoc hybrid between normal diffusion theory (which indeed has a prediction for the RMS path \cite{Dav02}) and its anomalous counterpart (which still has none).  This modification of the basic model can be viewed as significant, meaning that we are in principle back to validation step 1 with the new model.  However, this exercise uncovered something quite telling about the original anomalous diffusion model, namely, that its simple asymptotic (large optical depth) form used in all the above tests is not generally valid: for typical cloud covers, the pre-asymptotic terms computed explicitly for the normal diffusion case prove to be important irrespective of whether the diffusion is normal or not. Consequently, in its original form (resulting in a simple scaling law for the mean path with respect to cloud thickness and optical depth), the anomalous diffusion model fails to reproduce the new data even for the mean path. (Consequently, previous fits yielded only ``effective'' anomaly parameters and were misleading if taken too literally.) So we assign $p/q=0.1$ at best for the original model, hence $F^{(5)}=4\,10^{-4}$.

\end{enumerate}

\noindent 
Thus, $V_{\rm posterior}^{(6)} / V_{\rm prior}^{(1)} = 3\,10^{-3}$, a fatal blow for the anomalous diffusion in its simple asymptotic form, even though $V_{\rm posterior}^{(5)} / V_{\rm prior}^{(1)} = 7.0$ which would have been interpreted as close to a convincing validation.

This is not the end of the story, of course.  The original model has already spawned Scholl et al.'s empirical hybrid and a formalism based on integral (in fact, pseudo-differential) operators has been proposed \cite{Bul01} that extends the anomalous {\it diffusion} model to pre-asymptotic regimes. More recently, a model for anomalous {\it transport} (i.e., where angular details matter) has been proposed that fits all of the new oxygen spectroscopy results \cite{Dav06}.

In summary, the first and simplest incarnation of the anomalous diffusion model for solar photon transport ran its course and demonstrated the power of oxygen-line spectroscopy as a test for the performance of radiative transfer models required in climate modeling for large-scale average responses to solar illumination. Eventually, new and interesting tests will become feasible when we obtain dedicated oxygen-line spectroscopy from space with NASA's Orbiting Carbon Observatory (OCO) mission planned for launch in 2008. Indeed, we already know that the asymptotic scaling for reflected photon paths \cite{Dav99} is different from their transmitted counterparts \cite{Dav02} in standard diffusion theory for both mean and RMS.  

\subsection{A Computational Fluid Dynamics Model for Shock-Induced Mixing}

\begin{figure}
\begin{center}
\includegraphics[width=6cm]{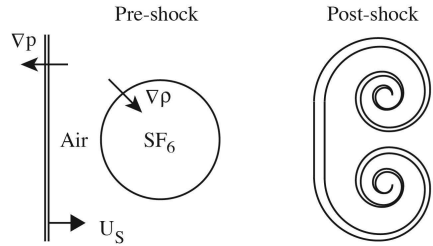}
\caption{
Schematic of the interactions between weakly shocked (Mach number $\approx$1.2) light gas (air) and a column of dense gas (SF$_6$). The Richtmyer--Meshkov instability occurs from the mismatch between the pressure gradient (at the shock front) and the density gradient (between the light and dense gases), which acts as a source of baroclinic vorticity. The column of dense gas ``rolls up'' into a double-spiral form under the action of the evolving vorticity.
}
\label{ShockTubeFig_Kamm_new}
\end{center}
\end{figure}

So far, our examples of models for complex phenomena have hailed from quantum and statistical physics.  In the latter case, they are stochastic models composed of: (1) simple code (hence rather trivial verification procedures) to generate realizations, and (2) analytical expressions for the ensemble-average properties (that are used in the above validation exercises). We now turn to gas dynamics codes which have a broad range of applications, from astrophysical and geophysical flow simulation to the design and performance analysis of engineering systems. Specifically, we discuss the validation of the ``Cuervo'' code developed at  Los Alamos National Laboratory \cite{Rid05,Rid07} for use as a simulation tool in the complex physics of compressible mixing. This software generates solutions of the Euler equations for flows of inviscid, non-heat-conducting, compressible gas. Cuervo has been verified against a suite of test problems including, e.g., those discussed by Liska and Wendroff \cite{Lis03}. As clearly stated by Oberkampf and Trucano \cite{Obe02a} however, such verification differs from and does not guarantee validation against experimental data. A standard validation scenario involves the Richtmyer--Meshkov (RM) instability \cite{Ric60,Mes69}, which arises when a density gradient in a fluid is subjected to an impulsive acceleration, e.g., due to passage of a shock wave (see Fig.~\ref{ShockTubeFig_Kamm_new}). Evolution of the RM instability is nonlinear and hydrodynamically complex and hence defines an excellent problem-space to assess CFD code performance for more general mixing scenarios.

In the series of shock-tube experiments described in \cite{Ben04}, RM dynamics are realized by preparing one or more cylinders with approximately identical axisymmetric Gaussian concentration profiles of dense sulfur hexaflouride (SF$_6$) in air. This (or these) vertical ``gas cylinder(s)'' is (are) subjected to a weak shock---Mach number $\approx$1.2---propagating horizontally, i.e., perpendicular to the axis of the gas cylinders. The ensuing dynamics are largely governed by the mismatch of the density gradient between the gases (with the density of SF$_6$ approximately five times that of air) and the pressure gradient through the shock wave;  this mismatch acts as the source for baroclinic vorticity generation. Moreover, the flow evolution is strongly two-dimensional up to the final times considered. Visualization of the density field is obtained using a planar laser-induced fluorescence (PLIF) technique, which provides high-resolution quantitative concentration measurements in a plane that cross-cuts the cylinders. The velocity field is diagnosed using particle image velocimetry (PIV), based on correlation measurements of small-scale particles that are seeded in the initial flow field. Careful post-processing of images from 130~$\mu$s to 1000~$\mu$s after shock passage yields planar concentration and velocity with error bars.

\begin{enumerate}

\item This RM flow is dominated at early times by a vortex pair. Later, secondary instabilities rapidly transition the flow to a mixed state. We rate $c_{\rm novel}=10$ for the observations of these two instabilities. The Cuervo code correctly captures these two instabilities, best observed and modeled with a single cylinder. At this qualitative level, we rate $p/q=10$ (good fit), which leads to $F^{(1)}=2.4$.

\item Older data for two-cylinder experiments acquired with a fog-based technique (rather than PLIF) showed two separated spirals associated with the primary instability, but the Cuervo code predicted the existence of a material bridge between those structures. This previously unobserved connection was subsequently diagnosed experimentally with the improved observational technique, i.e., the simulation code was truly predictive of this phenomenon. Using $c_{\rm novel}=10$ and $p/q=10$ yields $F^{(2)}=2.4$.

\item The evolution of the total power as a function of time offers another useful metric. The numerical simulation quantitatively accounts for the exponential growth of the power with time, within the experimental error bars. Using $c_{\rm novel}=10$ and $p/q=10$ yields $F^{(3)}=2.4$.

\item The concentration power spectrum as a function of wavenumber for different times provides another way (in the Fourier domain) to present the information of the hierarchy of structures already visualized in physical space ($c_{\rm novel}=1$). The Cuervo code correctly accounts for the low wavenumber part of the spectrum but underestimates the high wavenumber part (beyond the deterministic-stochastic transition wavenumber) by a factor 2 to 5. We capture this by setting $p/q=0.1$, which yields $F^{(4)}=0.47$.

\end{enumerate}

\noindent 
Combining the multipliers according to (\ref{mglerdffd}) leads to $V_{\rm posterior}^{(5)} / V_{\rm prior}^{(1)} = 6.5$, a significant gain, but still not sufficient to compellingly validate the Cuervo code for inviscid shock-induced hydrodynamic instability simulations, at least in 2D. Clearly, validation against this single set of experiments is inadequate to address all intended uses of a CFD code such as Cuervo.
\footnote{
Intricate experiments with three gas cylinders have since been performed \cite{Kum05} and others are currently under way to further challenge compressible flow codes.
}

\subsection{Discussion}

The above three examples illustrate the utility of representing the validation process as a succession of steps, each of them characterized by the two parameters $c_{\rm novel}$ and $p/q$. The determination of $c_{\rm novel}$ requires expert judgment and that of $p/q$ a careful statistical analysis, which is beyond the scope of the present report (see Ref.~\cite{Obe05} for a detailed case study). The parameter $q$ is ideally imposed as a confidence level, say $95\%$ or $99\%$ as in standard statistical tests. In practice, it may depend on the experimental test and requires a case-by-case examination.

The uncertainties of $c_{\rm novel}$ and of $p/q$ need to be assessed. Indeed, different statistical estimations or metrics may yield different $p/q$'s and different experts will likely rate differently the novelty $c_{\rm novel}$ of a new test. As a result, the trust gain $V_{\rm posterior}^{(n+1)} / V_{\rm prior}^{(1)}$ after $n$ tests necessarily has a range of possible values that grows geometrically with $n$. In certain cases, a drastic difference can be obtained by a change of $c_{\rm novel}$. For instance, if instead of attributing $c_{\rm novel} =100$ to the sixth OFC test, we put $c_{\rm novel}=10$ (resp. $1$) while keeping $p/q=0.1$, $F^{(6)}$ is changed from $4 \cdot 10^{-4}$ to $4 \cdot 10^{-3}$ (resp. $0.47$). The trust gain then becomes $V_{\rm posterior}^{(7)} / V_{\rm prior}^{(1)} = 0.07$ (resp. $\simeq 9$). For the sixth OFC test, $c_{\rm novel}=1$ is arguably unrealistic, given the importance of faults in seismology. The two possible choices $c_{\rm novel} =100$  and $c_{\rm novel} =10$ then give similar conclusions on the invalidation of the OFC model. In our examples, $V_{\rm posterior}^{(n+1)} / V_{\rm prior}^{(1)}$ provides a qualitatively robust measure of the gain in trust after $n$ steps; this robustness has been built-in by imposing a coarse-grained quality to $p/q$ and $c_{\rm novel}$.

\section{Summary}

The validation of numerical simulations continues to become more important as computational power grows, as the complexity of modeled systems increases, and as increasingly important decisions are influenced by computational models. We have proposed an iterative, constructive approach to validation using \emph{quantitative measures} and \emph{expert knowledge} to assess the relative state of validation of a model instantiated in a computer code.  In this approach, the increase/decrease in validation is mediated through a function that incorporates the results of the model vis-\`a-vis the experiment together with a measure of the impact of that experiment on the validation process. While this function is not uniquely specified, it is not arbitrary: certain asymptotic trends, consistent with heuristically plausible behavior, must be observed. In four fundamentally different examples, we have illustrated how this approach might apply to a validation process for physics or engineering models. We believe that the multiplicative decomposition of trust gains or losses (given in Eq. \ref{mglerdffd}), using a suitable functional prescription (such as Eq. \ref{mgler222}), provides a reasoned and principled description of the key elements---and fundamental limitations---of validation. It should be equally applicable to biological and social sciences, especially since it is built upon the decision-making processes of the latter. We believe that our procedure transforms the paralyzing criticisms in Popper's style that ``we cannot validate, we can only invalidate'' \cite{Ore94a} into a practical constructive algorithm. This strategy addresses specifically both problems of distinguishing between competing models and transforming the vicious circle of lack of suitable data into a virtuous spiral path: each cycle is marked by a quantified increment of the evolving trust we put in a model based on the novelty and relevance of new data and the quality of fits.

We have also surveyed and commented extensively on the V\&V literature. We hope this digest will help the reader as much as its collation helped us deepen our understanding of the challenge of model validation, including a new perspective on some of our own work.  We close with these far-reaching thoughts by Patrick J. Roache \cite{Roa04}:

\begin{quote}

{\it In an age of spreading pseudoscience and anti-rationalism, it behooves those of us who believe in the good of science and engineering to be above reproach whenever possible. Public confidence is further eroded with every error we make. Although many of society's problems can be solved with a simple change of values, major issues such as radioactive waste disposal and environmental modeling require technological solutions that necessarily involve computational physics. As Robert Laughlin~\cite{Lau02} noted in this magazine,} ``there is a serious danger of this power [of simulations] being misused, either by accident or through deliberate deception.'' {\it Our intellectual and moral traditions will be served well by conscientious attention to verification of codes, verification of calculations, and validation, including the attention given to building new codes or modifying existing codes with specific features that enable these activities.}

\end{quote}


\section*{Acknowledgments}

This work was supported by the LDRD 20030037DR project ``Physics-Based Analysis of Dynamic Experimentation and Simulation'' and the US DOE Atmospheric Radiation Measurement (ARM) program.  We acknowledge stimulating interactions and discussions with the other members of the project, including Bob Benjamin, Mike Cannon, Karen Fisher, Andy Fraser, Sanjay Kumar, Vladilen Pisarenko, Kathy Prestridge, Bill Rider, Chris Tomkins, Kevin Vixie, Peter Vorobieff, and Cindy Zoldi. We were particularly impressed by Timothy G. Trucano as our reviewer for \cite{Sor07}, who provided an extremely insightful and helpful report, with many thoughtful comments and constructive suggestions.  As authors, we count ourselves very fortunate to have had such a strong audience to scrutinize and improve our contribution.

\setcounter{equation}{0}
\numberwithin{equation}{section}

\appendix

\section*{Appendix: \\ 
A More Formal Look at the Role of Validation in the \\ 
Modeling Enterprise}

\setcounter{section}{1}

We deal with models that possess two aspects: a conceptual part based on the physical laws of nature (such as the Navier--Stokes conservation equations for fluid dynamics) and a computational part (like in CFD). Mathematically, a model along with observations are defined formally, as described in section \ref{mgmle} below:
\begin{itemize}
\item 
The model ${\cal M}$ maps the set $\{A\}$ of parameters and of initial and boundary conditions to a forecast of state variables in a formal vector $X_{\rm f}$;
\item 
An observation projection ${\cal G}$ maps the true dynamics or physics in $X_{\rm t}$ to raw measurements $y_{\rm o}$.
\end{itemize}
Such definitions may seem abstract and of little use but they are important foundations to build a comprehensive roadmap for physically-based model validation.

In the following section, we refine the above definitions and introduce a few more operators and quantities.  In section~\ref{s:4pbs}, we revisit the key steps in a validation loop with this notation in hand. Finally, we discuss some fundamental limitations on model validation in section~\ref{s:LimVal} using some of our own research in time-series analysis for illustration.

\subsection{Definitions}
\label{mgmle}

Let us denote $X_{\rm t}({\vec r},t)$ the true physical field. Observations $y_{\rm o}({\vec r},t)$ are obtained via a possibly nonlinear operator ${\cal G}$ acting on $X_{\rm t}({\vec r},t)$:
\begin{equation}
y_{\rm o}({\vec r},t) = {\cal G}\{ X_{\rm t}({\vec r'},t')\}~.
\label{eqa}
\end{equation}
The observations at position ${\vec r}$ and time $t$ may be a combination of past values obtained over some finite region, hence our use of $({\vec r'},t')$ which are different from $({\vec r},t)$.  The operator ${\cal G}$ may thus be non-local and (causally) time-dependent. In addition, any measurement has noise and uncertainties. Therefore, ${\cal G}$ is a stochastic operator. The simplest specification beyond ignoring noise is to consider an additive noise.

A model ${\cal M}$ provides a forecast $X_{\rm f}({\vec r},t)$ either in the actual future or in terms of what will lead (via another operator) to the value of the measurements beyond a certain fiducial point in time.  This is expressed by
\begin{equation}
X_{\rm f}({\vec r},t) = {\cal M}\left( \{ A \} \right)~.
\label{mjge}
\end{equation}
${\cal M}$ is the model operator, which contains for instance the equation of states, the formulation in terms of ODEs, PDEs, discrete maps and so on, which are supposed to embody the known physics of the underlying processes. $\{ A \}$ contains the parameters of the model as well as the boundary and initial conditions. The model operator ${\cal M}$ has a non-random part. It can also contain an additive or multiplicative noise component to represent the forecast errors as well as possible intrinsic stochastic components of the dynamics. The forecast errors may stem from computational errors, numerical instabilities and uncertainties, the existence of multiple branches in the solution and so on. The simplest specification is again to consider an additive noise.

The output ${\cal M}\left( \{ A \} \right)$ of the model is translated into physical quantities that can be compared with the observation via another operator ${\cal H}$, which models mathematically and in code the observation process.  In general, one would like to compare $y_{\rm o}({\vec r},t)$ given by (\ref{eqa}) with ${\cal H}\left[X_{\rm f}({\vec r},t) \right]$, that is,
${\cal G}\{ X_{\rm t}({\vec r'},t') \}$
with
${\cal H}\left[{\cal M}\left( \{ A \} \right)\right]$.
The intended use of the model is key to ``objective model validation,'' because it turns ``subjectiveness'' of the model validation into an ``object'' using hypothesis testing and decision theory. To implement this idea, it is natural to introduce a cost function (see below) for the intended use of the model:
\begin{equation*}
{\cal C}\left( {\cal G}\{ X_{\rm t}({\vec r'},t') \};
               {\cal H}\left[{\cal M}\left( \{ A \} \right)\right] \right)~,
\end{equation*}
which is a measure of how well the model accounts for the observations. In this expression, the cost function is evaluated in the ``physical space'' of observations/measurements. An alternative is to evaluate the cost function in the ``model space,'' i.e.,
\begin{equation*}
{\cal C}\left( {\cal G}^{-1}\{ {\cal H}\{ X_{\rm t}({\vec r'},t')\} \};
               {\cal M}\left( \{ A \} \right) \right)~,
\end{equation*}
where ${\cal G}^{-1}$ is the formal inverse operator to ${\cal G}$ which maps observations $y_{\rm o}$ onto the model space $X_{\rm f}$. In data assimilation, explicit form of ${\cal G}^{-1}$ does not exist in general due to rank deficiency. However, such alternative representation  within the linear theory corresponds to the duality between Kalman filtering and the 3D-Var \cite{Ide97}.

We propose to define the validation problem as a decision problem in which one uses the loss function to infer/decide how much confidence one feels in the reliability of the model to function in the range in which it is supposed to apply. The interesting and challenging situation occurs when this range extends beyond the region of parameter space 
in which all reasonably stringent controls have been performed. Validation requires the build-up of trust in the model or code so that it is believed to be resilient and to work in complex real situations combining the simple regimes that have been tested. The cost function is just an alternative way of constructing the statistical test that provides the probability level $p$ defined in the main text.

\subsection{Four Recurring Types of Problem in Physically-Based Model Validation}
\label{s:4pbs}

Our overarching goal is to advocate approaches to validation that are grounded in physics.  The term ``physics-based'' embodies two strategies:
\begin{itemize}
\item[(a)] 
use physical reasoning to improve modeling, target experiments and loss functions, and detect missed ``dimensions;''
\item[(b)] 
use concepts from statistical physics to formulate (in the spirit of Brown and Sethna \cite{Bro03}) a validation process of complex models with complex data in the form of an $N$-body problem.
\end{itemize}
Following this roadmap, we find ourselves asking the same four questions again and again: 
\begin{enumerate}
\item {\it How to model?}\/ (the question of model construction)
\item {\it What to measure?}\/ (the question of estimating $c_{\rm novel}$ in the main text)
\item {\it How to measure it?}\/ (the question of choosing and estimating the cost function or ``metric'')
\item {\it How to interpret the results?}\/ (the question of estimating $p$ in the main text)
\end{enumerate}
We view these four defining questions as the crucial steps within the validation loop described in sections~2--4 of the main text.

\subsubsection{Problem 1: Targeting model development (How to model?)}
\label{pb0}
Our discussion so far may give the impression that the modeling step is ``homogeneous.'' It may actually be advantageous to develop a hierarchical modeling framework. In this respect, Oden et al. \cite{Ode03} proposed to use hierarchical modeling as a mathematical structure that can be useful in directing validation studies. In this construction, a class of models of events of interest is defined in which one identifies a ``fine'' model that possesses a level of sophistication high enough to adequately capture the event of interest with good accuracy. This model may be intractable, even computationally. Hierarchical modeling consists in identifying a family of coarse models that are solvable. Using the fine model output as a datum, the error in the solution of ever coarser models can be estimated and controlled, with the goal of obtaining a model best suited for the simulation goal at hand. The essential components of this program are the following \cite{Ode03}:
\begin{enumerate}
\item
Experimental data are collected to fully characterize the fine model.
\item
Quantities ${\cal G}(X)$ of interest are specified as the essential physical entity to be predicted in the simulation (for instance in the form of the probability of the predicted values of the quantity).
\item
The coarsest model is used to extract a preliminary estimate of ${\cal G}(X)$ and modeling and approximation errors are computed.
\item
If the estimated error exceeds the prescribed tolerance, the model is enhanced and the calculation is repeated until a model yielding results within the preset bounds is obtained.
\item
The truncation error of the perturbation expansion is estimated: if the total error exceeds a preset tolerance, the data set and the fine model definition must be updated; if not, the predicted ${\cal G}(X)$ and the probability that it will take on values in a given interval are produced as output.
\end{enumerate}

A concrete implementation of this program has been performed by Israeli and Goldenfeld \cite{Isr04}. Using elementary cellular automata as an example, Israeli and Goldenfeld show how to coarse-grain cellular automata in all categories of Wolfram's exhaustive classification \cite{Wol02}. The main discovery is that computationally irreducible physical processes can be predictable and even computationally reducible at a coarse-grained level of description. The resulting coarse-grained cellular automaton constructed with the coarse-graining procedure emulate the large-scale behavior of the original systems without accounting for small-scale details. These results remind us that it is advantageous to develop a view of complex physical processes at different scales, as the predictability may depend on the scale of observation.

A related approach has been discussed recently by Brown and Sethna \cite{Bro03}, who consider models defined in terms of a set of nonlinear ODEs applied to systems that have large numbers of poorly known parameters, simplified dynamics, and uncertain connectivity. They call models possessing these three features, ``sloppy models.'' Sloppy models characterize many other high-dimensional multi-parameter nonlinear models. Brown and Sethna propose to use the maximum likelihood method to frame the problem of parameter estimation and model validation in the form of statistical ensemble method. In our language, the problem boils down to a study of the cost function ${\cal C}$ and its stiff and soft directions determined from the eigenvalue problem of the Hessian of ${\cal C}$ (with respect to the parameters of the model). In practice, Brown and Sethna propose to estimate the Hessian of ${\cal C}$ in terms of the so-called ``Levenberg-Marquardt'' Hessian (thus called because of its use of that popular minimization algorithm); that quantity is defined simply as a sum of pairwise products of first-order derivatives of the residuals with respect to the model parameters. Stiff modes correspond to large eigenvalues. Similar to a decomposition in principal components, retaining the stiff modes allows one to get a more robust signature of the coarse-grained properties of the dynamics. This constitutes a concrete implementation of our Problem 4 below on ``targeting model errors.'' This procedure also addresses the problem of defining the operator ${\cal H}$ that selects the output of the model for comparison to the experimental data.

There is an interesting avenue for research here: rather than performing the principal component decomposition in one step, it may be advantageous to perform a series of sub-system analysis, or cluster analysis, retaining the stiff modes of each sub-system and then aggregating them at the next level of the hierarchy.

\subsubsection{Problem 2: Targeting the observations (What to measure?)}
\label{pb1}
{\it Objective:} Find ${\cal G}$ (and the associated ${\cal H}$) that reveals the most about model critical behavior.

The problem has been addressed specifically in these terms by Palmer et al. \cite{Pal98} to target adaptive observations to ``sensitive'' parts of the atmosphere. Targeting observations could be directed by the desire to get access to the most relevant information that is also the most reliable (e.g., contaminated by the smallest errors). It may be worth mentioning that targeting the observations depends not only on ${\cal G}$, but also ${\cal M}$, $\{A\}$, as well as ${\cal C}$ (along with its own parameters discussed below). The targeting of the observations is the problem of maximizing the coefficient $c_{\rm novel}$ introduced in the main text so that the new experiment/observation explores novel dimensions of the parameter and variable spaces of both the process and the model that can best reveal potential flaws that could compromise the important applications. In general, one targets observations by developing experiments that are thought to provide, in some sense, the most relevant tests of the physics.

Oberkampf and Trucano (2002) \cite{Obe02a} suggest that traditional experiments could generally be grouped into three categories:
\begin{enumerate}
\item 
experiments that are conducted primarily for the purpose of improving the fundamental understanding of some physical process;
\item 
experiments conducted primarily for constructing or improving mathematical models of fairly well-understood flows;
\item 
experiments that determine or improve the reliability, performance, or safety of components, subsystems, or complete systems.
\end{enumerate}
These authors argue that validation experiments constitute a fourth type of experiment: {\it ``A validation experiment is conducted for the primary purpose of determining the validity, or predictive accuracy, of a computational modeling and simulation capability. In other words, a validation experiment is designed, executed, and analyzed for the purpose of quantitatively determining the ability of a mathematical model and its embodiment in a computer code to simulate a well-characterized physical process.''} This leads them to propose the following guidelines:
\begin{itemize}
\item {\it Guideline \#1}: 
A validation experiment should be jointly designed by experimentalists, model developers, code developers, and code users working closely together throughout the program, from inception to documentation, with complete candor about the strengths and weaknesses of each approach.
\item {\it Guideline \#2}: 
A validation experiment should be designed to capture the essential physics of interest, including all relevant physical modeling data and initial and boundary conditions required by the code.
\item {\it Guideline \#3}: 
A validation experiment should strive to emphasize the inherent synergism between computational and experimental approaches.
\item {\it Guideline \#4}: 
Although the experimental design should be developed cooperatively, independence must be maintained in obtaining both the computational and experimental results.
\item {\it Guideline \#5}: 
A hierarchy of experimental measurements of increasing computational difficulty and specificity should be made, for example, from globally integrated quantities to local measurements.
\item {\it Guideline \#6}: 
The experimental design should be constructed to analyze and estimate the components of random (precision) and bias (systematic) experimental errors.
\end{itemize}

\subsubsection{Problem 3: Targeting the cost function (How to estimate the 
penalty on imperfect models and measurements using their discrepancies?)}
\label{pb2}
For given measurements or experiments, that is, for given ${\cal G}$, the problem is to find the optimal cost function ${\cal C}$ for the intended use of the model. The notion of optimality needs to be defined. It could capture a compromise between the following requirements:
\begin{itemize}
\item 
fit best the important features of the data (what is ``important'' may be decided on the basis of previous studies and understanding or other processes, or programmatic concerns);
\item 
minimize the extraction of spurious information from noise, which requires one to have a precise idea of the statistical properties of the noise (if such knowledge is not available, the cost function should take this into account).
\end{itemize}

The choice of the cost function involves the choice of how to look at the data. For instance, one may want to expand the measurements at multiple scales using wavelet decompositions and compare the prediction and observations scale by scale, or in terms of multifractal spectra of the physical fields estimated from these wavelet decomposition or from other methods. The general idea here is that, given complex observation fields, it is appropriate to ``project'' the data onto a variety of ``metrics'' designed to detect and characterize phenomena of particular interest. For instance, wavelet-based scaling properties can be used in the comparison between observations and model predictions; the question is then: How well is the model/code able to reproduce the salient multi-scale properties derived from the observations? The physics of turbulent fields and of complex systems have offered many such new tools to unfold complex fields according to different statistics. Each of these statistics provides a basis for a metric to compare observations with model predictions. Each such statistics thus leads to a cost function focusing on a particular feature of the process. These metrics are derived from the understanding that turbulent fields can be analyzed using them, revealing strong constraints in their organization (spatial structure and temporal evolution). These metrics can therefore be described as ``physics-based.''

Furthermore, the choice of the cost function should take into account that the diagnostics of the experiments may lead to spurious results \cite{Cal02}. For example, in laser-driven shock experiments, because the laser-induced fluorescence method illuminates the mixing zone with a planar sheet of light, this diagnostic can lead to aliasing of long-wavelength structures into short-wavelength features in the images, thus affecting the interpretation of observed small-scale structures in the mixing zone. Also, because of the dynamic limits on diagnostic resolution, the formation of small-scale structure cannot be completely determined. 

As emphasized  by Noam Chomsky in his own field of work \cite{Cho81}, the danger with the Popperian strategy \cite{Pop59} is that one might prematurely reject a theory based on ``falsification'' using data that are themselves poorly understood. For instance, lack of quality control for the experiments can result in premature rejection of the model. On these issues, Stein \cite{Ste02} discusses means for controlling and for understanding sample selection and variability, which can compromise conclusions drawn from validation tests.

The problem of the choice of the cost function ${\cal C}$ seems, however, to be of less importance than Problem 2 above and Problem 4 below. In fact, almost all classical results on the limit properties of efficiency of statistical inference are valid (and proved) for a whole general family of cost functions ${\cal C}(\cdot;\cdot)$ satisfying the following conditions (see, e.g., Ibragimov and Hasminskii \cite{Ibr81}):
\begin{itemize}
\item[(a)] 
${\cal C}(x,y) = c(|x-y|)$;
\item[(b)] 
$c(z)$ is a positive monotonically increasing function (including, e.g., power-law functions $|z|^q$, with $q>0)$;
\item[(c)] 
$c(z)$ should not increase too fast (its mean with respect to the Gaussian distribution must remain finite).
\end{itemize}
Thus, statistical limit theorems are proved for the whole class of different power-law cost functions (including the classic choice $q=2$).

As an example, it may be appropriate to consider the cost function in the following form. Let us assume we are interested in some functional
\begin{equation*}
Z(R,T|{\cal G}\{X_{\rm t}({\vec r},t)\}, {\vec r} \in D(R), t \leq T)
\end{equation*}
depending on the past true physical field $X_{\rm t}({\vec r},t)$ in some region $D(R)$. In this case, the cost function can be chosen as
\begin{equation}
{\cal C} \left(
Z\left[R,T|{\cal G}\{X_{\rm t}({\vec r},t)\}, {\vec r} \in D(R), t \leq T\right]
;
Z\left[R,T|{\cal H}\{X_{\rm f}({\vec r},t)\}, {\vec r} \in D(R), t \leq T\right] 
\right)
\label{mgelw}
\end{equation}
where ${\cal C}(\cdot;\cdot)$ is some function satisfying above conditions (a)--(c). The formulation (\ref{mgelw}) for ${\cal C}(\cdot;\cdot)$ should not only be a function of ${\cal G}$ and ${\cal M}$, but also of those parameters that correspond to our best guess for the uncertainties, errors and noise. Indeed, in most cases, we can never know real uncertainties, errors and noise in ${\cal G}$ and ${\cal M}$ (or even ${\cal H}$). Hence, we must parameterize them based on our best guess. In data assimilation (described in the main text in relation to model calibration and validation), the accuracy of such parameterization is known to influence the results significantly.

Generalizations to (\ref{mgelw}) allowing for different fields in the two sets of variables in ${\cal C}$ are needed for some problems, such as in validation of meteorological models. For instance, consider a model state vector $X$ (dimension is on the order of $10^6$) which is computed on a fixed spatial grid. In general, the locations of the observations are not on the computational grid (for example, consider measurements with weather balloons released from the surface). Thus, the observation $Y$ is a function of $X$, but is not an attempt to estimate $X$ itself. Hence, if the cost function is quadratic, it has the form $(Y-H(X))^{\rm T} O^{-1} (Y-H(X))$ where $H$ acts on the interpolation function to pick up the model variable at the grid points close to the observed location, and $O$ is related to the error covariance. Let us imagine a validation case using satellite infrared images for $Y$ and atmospheric radiative state for $X$. Observations are quasi-uniform in space at a given time; at each time, available observations and their quality (represented  by $O$) may change, however. In this case, the cost function must take into account the mapping between $X$ and $Y$ so that we have ${\cal C}(X,Y)=C(|H(X)-Y|)$ rather than ${\cal C}(X,Y)=C(|X-Y|)$; therefore $(Y-H(X))^{\rm T} O^{-1} (Y-H(X))$ when ${\cal C}$ is quadratic. In addition, for heterogeneous observations (satellite images, weather balloon measurements, airplane sampling, and so on), cost functions should take into account all these data into account such as 
\begin{equation*} 
{\cal C}(x,y) = {\cal C}_{\rm satellite}(x,y)
              + {\cal C}_{\rm balloon}(x,y)
              + {\cal C}_{\rm airplane}(x,y)
              + \cdots 
\end{equation*} 
and each ${\cal C}$ may have a complex idiosyncratic observation function $H$. See Courtier et al. \cite{Cou98} for a discussion on cost functions for atmospheric models and observation systems.

\subsubsection{Problem 4: Targeting model errors (How to interpret the results?)}
\label{pb3}
The problem here is to find the ``dimensions'' of the model that are missing, misrepresented or erroneous. The question of how to interpret the results thus leads to the discussion of the missing or misrepresented elements in the model.  What tests can be used to betray the existence of hidden degrees of freedom and/or dimensions?

This is the hardest problem of all. It can sometimes find an elegant solution when a given model is embedded in a more general model. Then, the limitation of the ``small'' model becomes clear from the vantage of the more general model. Well-known examples are
\begin{itemize}
\item 
Newtonian mechanics as part of special relativity, when $v \ll c$ where $v$ (resp. $c$) is the velocity of the body (resp. of light);
\item 
classical mechanics as part of quantum mechanics when $h/mc \ll L$ (where $h$ is Planck's constant, $m$ and $L$ are the mass and size of the body and
$h/mc$ is the associated Compton wavelength);
\item 
Eulerian hydrodynamics as part of Navier-Stokes hydrodynamics with its rich phenomenology of turbulent motion (when the Reynolds number goes to infinity, equivalently, viscosity goes to zero);
\item 
classical thermodynamics as part of statistical physics of $N \gg 1$ particles or elements, where phase transitions and thermodynamic phases emerge in the limit $N \to \infty$.
\end{itemize}
The challenge of targeting model errors is to develop diagnostics of missing dimensions even in absence of a more encompassing model. This could be done by adding random new dimensions to the model and studying its robustness.

In what sense can one detect that a model is missing some essential ingredient, some crucial mechanisms, or that the number of variables or dimensions is inadequate? To use a metaphor, this question is similar to asking ants living and walking on a plane to gain awareness that there is a third dimension. 
\footnote{
This question (raised already by the German philosopher Kant) actually has an answer that has been studied and solved by Ehrenfest in 1917 \cite{Ehr17} (see also Whitrow's 1956 article \cite{Whi56}). This answer is based on the analysis of several fundamental physical laws in $\mathbb{R}^n$ spaces and comparing their predictions as a function of $n$. The value $n=3$ turns out to be very special! Thus, ants studying gravitation or electro-magnetic fields will see that there is more to space than their plane.
}

\subsection{Fundamental Limits on Model Validation}
\label{s:LimVal}

Before, while and after engaging in model validation, it is wise to reflect frequently and systematically on what is not known. Two examples using the formalization introduced in section \ref{mgmle} are:

\subsubsection{Ignorance on the model ${\cal M}(\{A\})$}
As quoted in the main text, Roache \cite{Roa98a} states, in a nutshell, that validation is about solving the right equations for the problem of immediate concern. How do we know the right equations?

Consider, for instance, point vortex models, and let us perform ``twin experiments,'' i.e., (1) first generate the ``simulated observations'' by a ``true'' point vortex system that are unknown to the make-believe observer and modeler; (2) use the procedure of section \ref{mgmle} and construct a ``validated'' point vortex system. The problem is that, even before we start model validation, we are already using one of the most critical pieces of information, which is that the system is based on point vortices. Similar criticism for the use of ``simulated observations'' has been raised in data assimilation studies using OSSEs (Observing-System Simulation Experiments). This criticism is crucial for model validation.

For this unavoidable issue of model errors, we suggest that one needs a hierarchy of investigations:
\begin{enumerate}
\item 
Look at the statistical or global properties of the time series and/or fields generated by the models as well as from the data, such as distributions, correlation functions, $n$-point statistics, fractal and multifractal properties of the attractors and emergent structures, in order to characterize how much of the data our model fits. Part of this approach is the use of maximum likelihood theory to determine the most probable value of the parameters of the model, conditioned on the realization of the time series. 
\item 
We can bring to bear on the problem the modern methods of computational intelligence (or machine learning), including pattern classification and recognition methods ranging from the already classical ones (e.g., neural networks, $K$-means) to the most recent advances (e.g., support vector machines, ``random forests''). 
\item 
Lastly, a qualification of the model is obtained by testing and quantifying how well it predicts the ``future'' beyond the interval used for calibration/initialization.
\end{enumerate}

\subsubsection{Levels of ignorance on the observation ${\cal G}$}
\begin{itemize}
\item 
{\it First level}: The characteristics of the noise are known, such as its distribution, covariance, and maybe higher-order statistics.
\item 
{\it Second level}: It may happen that the statistical properties of the noise are poorly known or constrained.
\item 
{\it Third level}: A worse situation is when some noise components are not known to exist and are thus simply not considered in the treatment. For instance, imagine that one forgets in climate modeling about the impact of biological variability in time and space in the distribution of CO$_2$ sequestration sites.
\item 
{\it Fourth level}: Finally, there is the representation error in ${\cal G}$ itself, i.e., how ${\cal G}$ is modeled mathematically in ${\cal H}$.
\end{itemize}

\subsubsection{Consequences of the sensitivity to initial conditions and nonlinearity in the model}

Even an accurate forecast is limited by the inherent predictability of the system. In the same way, validation may be hindered by limited access to testing.
The predictability of a system refers to the fundamental limits of prediction for a system. For instance, if a system is pure noise, there is no possibility of forecasting it better than chance. Similarly, there may be limits in the possibilities of testing
the performance of a model because of limits in measurements, limits in access to key parameters for instance. With such limitations, it may be impossible to fully validate a model.

A well-known source that limits predictability is the property of sensitivity to initial conditions, which is one of the ingredients leading to chaotic behavior. Validation has to be made immune to this sensitivity upon initial conditions, by using a variety of methods, including the properties of attractors, their invariant measures, the properties of Lyapunov exponents, and so on. Pisarenko and Sornette \cite{Pis04} have shown that the sensitivity upon initial conditions leads to a limit of testability
in simple toy models of chaotic dynamical systems, such as the logistic map. They addressed the possibility of applying standard statistical methods (the least square method, the maximum likelihood estimation method, the method of statistical moments for estimation of parameters) to deterministically chaotic low-dimensional dynamic system containing an additive dynamical noise. First, the nature of the system is found to require that any statistical method of estimation combines the estimation of the structural parameter with the estimation of the initial value. This is potentially an important lesson for such a class of systems. In addition, in such systems, one needs a trade-off between the need of using a large number of data points in the statistical estimation method to decrease the bias (i.e., to guarantee the consistency of the estimation) and the unstable nature of dynamical trajectories with exponentially fast loss of memory of the initial condition. In this simple example, the limit of testability 
is reflected in the absence of theorems on the consistency and efficiency of maximum likelihood estimation (MLE) methods \cite{Pis04}. We can use MLE with sometimes good practical results in controlled situations for which past experience has been accumulated but there is no guarantee that the MLE will not go astray in some cases.

This work has also shown that the Bayesian approach to parameter estimation of chaotic deterministic systems is incorrect and probably suboptimal. The Bayesian approach usually assumes non-informative priors for the structural parameters of the model, for the initial value and for the standard deviation of the noise. This approach turns out to be incorrect, because it amounts to assuming a stochastic model, thus referring to quite another problem, since the correct model is fundamentally deterministic (only with the addition of some noise).

This negative conclusion on the use of the Bayesian approach should be contrasted with the Bayesian approach of Hanson and Hemez \cite{Han03} to model the plastic-flow characteristics of a high-strength steel by combining data from basic material tests. The use of a Bayesian approach to this later problem seems warranted because the priors reflect the intrinsic heterogeneity of the samples and the large dispersion of the experiments. In this particular problem concerning material properties, the use of Bayesian priors is warranted by the fact that the structural parameters of the model can be viewed as drawn from a population. It is very important to stress this point: Bayesian approaches to structural parameter determination are justified only in problems with random distributions of the parameters. For the previous problem of deterministic nonlinear dynamics, it turns out to be fundamentally incorrect. We therefore view proper partition of the problem at hand between deterministic and random components as an essential part of validation.

\subsubsection{Extrapolating beyond the range of available data}

In the previous discussion, the limit of testability 
is solely due to the phenomenon of sensitive dependence upon initial conditions, as the model is assumed to be known (the logistic map in the above example). In general, we do not have such luxury. Let us illustrate the much more difficult problem by two examples stressing the possibility for the existence of ``indistinguishable states.'' Consider a map $f_1$ that generates a time series. Assuming that $f_2$ is unknown a priori, let us construct/constrain the map $f_2$ whose initial condition and parameters can be tuned in such a way that trajectories of $f_2$ can follow data of $f_1$ for a while, but eventually the two maps diverge. Suppose that the time series of $f_1$ is too short to explore the range expressing the divergence between the two maps. How can we (in-)validate $f_2$ as a incorrect model of $f_1$?

This problem arises in the characterization of the tail of distributions of stochastic variables. For instance, Malevergne, Pisarenko and Sornette \cite{Mal05} have shown that, based on available data, the best tests and efforts can not distinguish between a power law tail and a stretched exponential distribution for financial returns. The two classes of models are indistinguishable, given the amount of data. This fundamental limitation has unfortunately severe consequences, because choosing one or the other models involves different predictions for the frequency of very large losses that lie beyond the range sampled by historical data (the $f_1-f_2$ problem). The practical consequences are significant, in terms of the billions of dollars banks should put (or not) aside to cover large market swings that are outside the data set available from the known past history.

This example illustrates a crucial aspect of model validation, namely that it requires the issuance of predictions outside the domain of parameters and/or of variables that has been tested ``in-sample'' to establish the (calibrated or ``tuned'') model itself.


\bibliographystyle{unsrt}

\bibliography{CMT-07_paper}

\printindex

\end{document}